\documentclass[journal]{IEEEtran}
\usepackage{amsmath}
\usepackage{amssymb}
\usepackage{algorithm}
\usepackage{algorithmicx}
\usepackage{algpseudocode}
\usepackage[T1]{fontenc}
\usepackage{color}
\usepackage{times}
\usepackage{cite}
\usepackage{caption}
\usepackage{graphicx}
\usepackage{float} 
\usepackage{subcaption}
\usepackage{setspace}%行距
\usepackage{stfloats}

\begin{document}
	\newcounter{TempEqCnt}
\title{\huge Enhancing Spectrum Sensing via Reconfigurable Intelligent Surfaces:  Passive or Active Sensing and How Many Reflecting Elements are Needed?}
\author{Hao Xie, Dong Li,~\IEEEmembership{Senior Member,~IEEE,} and Bowen Gu %\vspace{-20pt}
	\thanks{H. Xie, D. Li, and B. Gu are with the School of Computer Science and Engineering, Macau University of Science and Technology, Avenida Wai Long, Taipa, Macau 999078, China (e-mails: 3220005631@student.must.edu.mo, dli@must.edu.mo, 21098538ii30001@student.must.edu.mo).}
%	\thanks{Z. Lin is with the College of Electronic Engineering, National University of Defense Technology, Hefei 230037, China, is also with the School of Computer Science and Engineering, Macau University of Science and Technology, Taipa, Macau 999078, China (e-mail: linzhi945@163.com).}	
%	\thanks{Y. Xu is with the School of Communication and Information Engineering, Chongqing University of Posts and Telecommunications, Chongqing 400065, China (e-mail: xuyj@cqupt.edu.cn).}
} % <-this % stops a space
%, 

\maketitle
\begin{abstract}
Cognitive radio has been proposed to alleviate the scarcity of available spectrum caused by the significant demand for wideband services and the fragmentation of spectrum resources. However, sensing performance is quite poor due to the low sensing signal-to-noise ratio, especially in complex environments with severe channel fading. Fortunately, reconfigurable intelligent surface (RIS)-aided spectrum sensing can effectively tackle the above challenge due to its high array gain. Nevertheless, {\color{black}the traditional passive RIS may suffer from the ``double fading'' effect, which severely limits the performance of passive RIS-aided spectrum sensing}. Thus, a crucial challenge is how to fully exploit the potential advantages of the RIS and further improve the sensing performance. To this end, we introduce the active RIS into spectrum sensing and respectively formulate two optimization problems for the passive RIS and the active RIS to maximize the detection probability. In light of the intractability of the formulated problems, we develop a one-stage optimization algorithm with inner approximation and a two-stage optimization algorithm with a bisection method to obtain sub-optimal solutions, and apply the Rayleigh quotient to obtain the upper and lower bounds of the detection probability. Furthermore, in order to gain more insight into the impact of the RIS on spectrum sensing, we respectively investigate the number configuration for passive RIS and active RIS and analyze how many reflecting elements are needed to achieve the detection probability close to 1. Simulation results verify that the proposed algorithms outperform existing algorithms under the same parameter configuration, and achieve a detection probability close to 1 with even fewer reflecting elements or antennas than existing schemes.
\end{abstract}

%
%
%
%
%
%
%
%
%
%\vspace{-10pt}
\section{Introduction}
\subsection{Backgroud}
%	The rapid development of Internet of Things (IoT) technologies has facilitated the application and connectivity of various IoT devices such as passive tags, industrial sensors, and controllers, which will create a highly interconnected ecosystem for various domains\cite{a1,a2}.  However, these IoT devices and sensors require frequent wireless communication and data exchange to achieve real-time data collection, analysis, intelligent decision-making, and response, which exerts pressure on the limited spectrum resources since each IoT device requires a portion of the available spectrum to transmit and receive data.  Furthermore, different IoT applications operate on different frequency bands, and thus the diversity of IoT applications such as environmental monitoring and position tracking will lead to the fragmentation of spectrum resources, which further aggravates the scarcity of available spectrum. Therefore, how to ensure the rational utilization and equitable allocation of spectrum resources is quite significant.
	
	\color{black}The modern digital society is currently experiencing a constant surge in wireless traffic, which may lead to the fragmentation of spectrum resources and further aggravate the scarcity of available spectrum\cite{a1}. \color{black}  Cognitive radio is a revolutionary wireless communication technology aiming at improving the utilization of spectrum resources\cite{b1}. {\color{black}This is different from the traditional static spectrum allocation schemes}, which have demonstrated poor performance in available spectrum utilization. In cognitive radio communication, secondary users (SUs) can access the spectrum owned by the primary network via the spectrum underlay mode or the spectrum overlay mode\cite{b2}. In the spectrum underlay mode, primary users (PUs) can share the spectrum with SUs simultaneously while the interference from the secondary network to the primary network must be maintained at an interference temperature level\cite{b3}. However, {\color{black}in unfavorable wireless transmission environments},  the spectrum underlay mode may cause higher interference to the PUs such that their quality of service requirements may be violated. Moreover, {\color{black}since the secondary network  needs to limit the interference power, it leads to severe performance degradation of the secondary network.} For the spectrum overlay mode, cognitive radio has the capability to monitor and analyze the surrounding spectrum environment in a real-time manner {\color{black}by applying spectrum sensing technologies}\cite{b4}. By collecting and analyzing wireless signals, cognitive radio can detect the spectrum occupied by PUs  and identify the idle spectrum. Once cognitive radio detects that a frequency band is in an idle state, secondary networks can autonomously utilize that band for communication, thus avoiding causing interference to PUs, which can effectively alleviate spectrum scarcity for 5G networks. 
	
	To achieve spectrum sensing with a high sensing performance, all kinds of spectrum sensing algorithms have been proposed, such as energy detection\cite{c1}, the matched filtering  detection\cite{c2}, the cyclostationary detection\cite{c3}, and the eigenvalue-based detection\cite{c4}. Spectrum sensing typically requires accurate detection of spectrum occupancy in low signal-to-noise ratio (SNR) environments. However, this can be hindered by factors such as channel fading, multi-path effects, and noise. In this case, limited spectrum sensing accuracy makes it difficult for the secondary transmitter (ST) to accurately detect spectrum occupancy, resulting in low spectrum utilization. There have been some methods to overcome the above challenges. First, introducing a collaborative sensing mechanism in cognitive radio is one of the feasible methods. However, the method requires devices to remain synchronized, {\color{black}which may be affected by factors such as clock drift and transmission delays so that the accuracy and timeliness of the detection may be decreased.}
	%然而,实现协作感知需要设备之间进行通信和信息交换，这会增加通信开销和能量消耗。协作感知需要设备之间保持同步，以便正确地协调和共享感知结果。同步要求可能会受到时钟漂移、传输延迟和网络拓扑等因素的影响，导致感知的不准确性和时效性下降。
	Second,  by combining multiple sensing algorithms, the detection probability can be improved, which, however, increases the complexity of the system, and it is also challenging to determine the optimal combination. 
	Third, increasing the sensing time can also improve sensing performance. However, longer sensing time indeed achieves better detection {\color{black}performance, which leads to fewer time resources for wireless transmission in secondary networks and thus greatly limits} the performance of the secondary networks. Therefore, more effective technologies are urgently needed to improve the sensing performance.
	
	Reconfigurable intelligent surface (RIS)\footnote{\color{black}Passive RIS is designed to passively reflect incident signals without actively generating or amplifying the signal. It may consume a small amount of energy for phase adjustment, but this consumption is practically negligible as compared to the much higher transmit power of active devices\cite{d0}.}, as an emerging wireless communication technology, has attracted wide attention and research due to its high array gain, low cost, and low power\cite{d0,d1}. The RIS utilizes reflecting elements with adjustable phase and amplitude to precisely control and manipulate wireless signals. {\color{black}Introducing RIS in the signal propagation path can alter the direction and phase of the signals}, thereby achieving signal enhancement, focusing, and directional transmission. {\color{black}Most existing works have applied} the RIS to the wireless information transfer\cite{d2,d3,d4} and the wireless energy transfer\cite{d5,d6,d7}. However, the enormous benefits of the RIS to spectrum sensing have not been fully exploited. In particular, the RIS, by adjusting the phase shift, can enhance and focus the wireless signal transmitted by the primary transmitter (PT), thereby introducing reflecting links to overcome the signal attenuation and interference issues caused by fading and blocking in traditional wireless communications. So far, little attention has been paid to RIS-aided spectrum sensing\cite{e1,e2,e3,e4,e5}.
%	In particular,  the author proposed a weighted energy detection method, and derived the closed-form expression of detection probability. The authors in [] also obtain the closed-form expressions of detection probability for 
	These works primarily analyzed closed-form expressions of detection probability from the perspective of a single RIS\cite{e1} or multiple RISs\cite{e2}. By optimizing the phase-shift matrix, the detection probability can be improved\cite{e3}. In addition to the energy detection algorithm, the authors in \cite{e4} also employed the maximum eigenvalue detection method and analyzed the number of reflecting elements required to achieve a detection probability close to 1. Furthermore, the authors in \cite{e5} also combined sensing with communication and utilized optimization theory to enhance the throughput of the secondary network.
	
	\vspace{-15pt}
\subsection{Motivation and Contributions}
\color{black}Although some works have been conducted on RIS-aided spectrum sensing in recent years, some fundamental issues still remain unsolved. First, these works primarily focus on performance analysis\cite{e1,e2,e4}, such as false alarm probability, or on enhancing the transmission performance of secondary networks\cite{e5}, which may not effectively improve sensing performance. Second, undoubtedly, the passive RIS can address signal attenuation or blockage issues in traditional wireless communications so that the detection probability can be improved. However, there are still some design issues that have not been taken into account, which means the advantages of the RIS can not be perfectly presented, i.e., the passive RIS-aided spectrum sensing will suffer from the ``double fading'' effect\footnote{\color{black}The “double fading” effect is also called product-distance path loss since the path loss of the cascaded channel is the product of the path losses of the transmitter-RIS and the RIS-receiver links.} \cite{i5} since the signal undergoes cascaded channels, i.e., from the transmitter to the RIS and from the RIS to the receiver, the path loss of the cascaded channel is much larger than that of the unobstructed direct channel, which has become a major bottleneck in restricting the performance of the RIS-aided spectrum sensing. Thus, a natural question arises: How to fully exploit the potential advantages of the RIS and further improve the sensing performance? \color{black}

{\color{black}Motivated by the above observations, we consider how to utilize passive/active RIS\cite{i5} for spectrum sensing to improve the sensing performance. The basic idea behind this active RIS\footnote{\color{black}Active RIS, while capable of amplifying received signals like amplify-and-forward relays, has significantly lower power consumption, hardware complexity, and hardware costs since it lacks the radio-frequency chains\cite{i5,i51}. Additionally, the active RIS inherits the adjustable phase-shift capability from the passive RIS, while the amplify-and-forward relay does not have this function.}\cite{i51} is that,} by allocating a certain power budget to the active RIS, the active RIS can adjust the phase shift and amplify the received signal simultaneously, which can mitigate signal attenuation and the ``double fading'' effect. The constructive thermal noise introduced by the active reflecting elements can enhance the sensing performance, which is completely different from the harmful nature of thermal noise in active RIS-aided wireless communications. Note that active RIS-aided spectrum sensing has received little attention, in which \cite{f1} is highly related to this work. However, \cite{f1} focused more on the trade-off between spectrum sensing and communication performance rather than sensing performance, which may not contribute much in terms of improving the sensing performance and is significantly in contrast to our work regarding the problem formulation and resulting solution methodology. In this paper, our goal is to maximize the sensing performance while {\color{black}ensuring a target probability of false alarm for the passive RIS and probabilistic power consumption for the active RIS.} Followed by the considered spectrum sensing algorithms, the resulting number configuration for the passive/active RIS, the upper and lower bounds of the detection probability, and the detection performance comparison of the two types (active and passive) of RIS are worth investigating, which are not presently available in existing works. The contributions of this work are summarized as follows:

\begin{itemize}
	\item {\color{black}We propose the passive/active RIS-aided spectrum sensing systems, where we formulate two different detection probability maximization problems for the passive/active RIS by jointly optimizing the receive beamforming vector, the phase-shift matrix, the detection threshold, and the amplification factor subject to the maximum power constraint of the RIS, the maximum probability of false alarm, and phase-shift constraint of the RIS.}
	\item {\color{black}In light of the intractability of the formulated problems, we respectively develop an alternating optimization (AO) method and a one-stage optimization algorithm with inner approximation for the passive/active RIS-aided spectrum sensing to obtain the sub-optimal solutions, where the closed-form solution of the detection threshold is also obtained. In order to further enhance the sensing performance for the active RIS, we propose a two-stage optimization algorithm, where the outer loop adopts a bisection method, and the inner loop employs an AO method. Besides, we obtain the upper and lower bounds of the detection probability by the Rayleigh quotient.}
	\item To gain more insight into the impact of the RIS on spectrum sensing, we investigate the number configuration for {\color{black}the passive/active RIS-aided spectrum sensing}. Specifically, we derive how many reflecting elements are required to achieve a detection probability close to 1 {\color{black}for the passive/active RIS}. Furthermore, we also compare the detection performance for the two types of RIS.
	\item Simulation results verify the superiority of the proposed algorithms. Compared to the algorithm with the passive RIS and without the RIS, the proposed algorithm {\color{black}with the active RIS} requires only a small number of reflecting elements or antennas to achieve a detection probability close to 1. By properly designing the number of reflecting elements and the maximum power of the active RIS, the proposed algorithms outperform the existing algorithms.
\end{itemize}
{\color{black}\subsection{Organization}
The remainder of this paper is structured as follows: The system models are presented in Section  \uppercase\expandafter{\romannumeral2}. The algorithm designs for the passive/active RIS-aided spectrum sensing are presented in Section \uppercase\expandafter{\romannumeral3} and \uppercase\expandafter{\romannumeral4}, respectively. Section \uppercase\expandafter{\romannumeral5} introduces the number configuration for the passive/active RIS-aided spectrum sensing.  Section \uppercase\expandafter{\romannumeral6} gives simulation results. The paper is concluded in Section \uppercase\expandafter{\romannumeral7}.}

\section{\color{black}System Models}
\begin{figure}%\vspace{-10pt}
	\centering
	\includegraphics[width=1.7in]{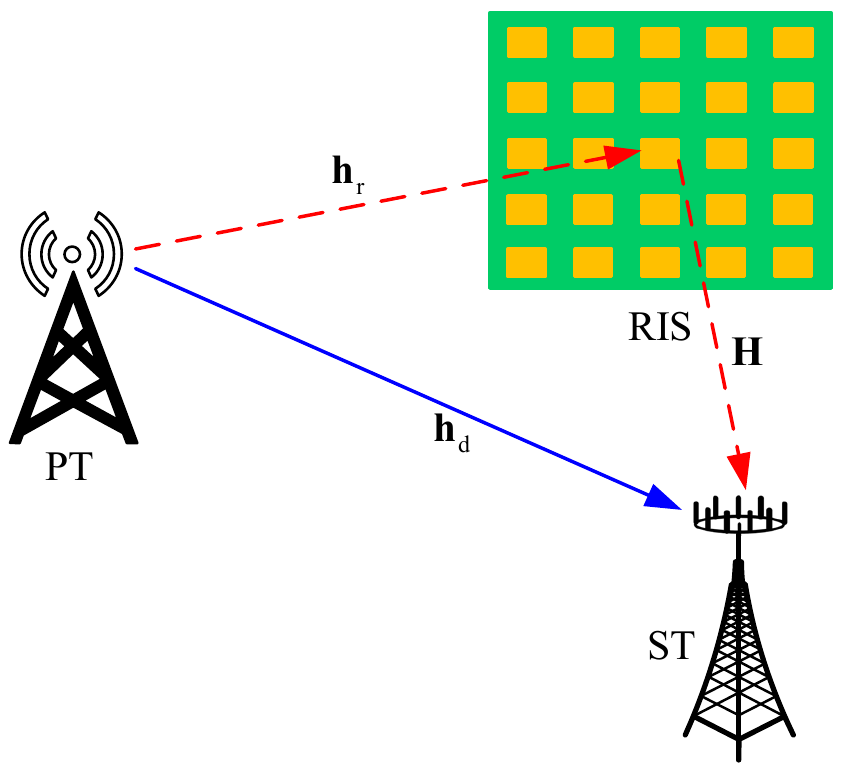}
	\caption{An RIS-enhanced spectrum sensing.}
	\label{fig0}%
\end{figure}

	As illustrated in Fig. \ref{fig0}, we consider an RIS-aided spectrum sensing\footnote{\color{black}For scenarios involving multiple PTs, the proposed algorithms can be directly applied since different primary networks can utilize orthogonal frequency bands. For scenarios involving multiple STs, we can also extend our sensing algorithms from our work into cooperative sensing algorithms by employing data fusion or decision fusion schemes\cite{c1}.} consisting of a PT with a single antenna, an ST with $M$ antennas\footnote{\color{black}In certain network scenarios with constraints such as power consumption, size, and cost, there may also exist configurations with the single-antenna PT and multi-antenna ST, such as some low-power devices, backscatter tags, and sensor networks. This configuration can reduce the complexity of the PT and enhance the sensing performance. Furthermore, by adjusting the dimension of transmit power, the proposed algorithms can be directly applied to the scenario with the multi-antenna PT.}, and an RIS with $N$ reflecting elements, where the ST opportunistically occupies the PT’s licensed spectrum, {\color{black}the RIS is deployed within the cell to assist in spectrum sensing}. Let us define $\boldsymbol{\rm h}_{\rm d}\in\mathbb{C}^{M\times 1}$, $\boldsymbol{\rm h}_{\rm r}\in\mathbb{C}^{N\times 1}$, and $\boldsymbol{\rm H}\in\mathbb{C}^{M\times N}$ as the channels from the PT to the ST, from the PT to the RIS, and from the RIS to the ST, respectively. 
\subsection{\color{black}System Model for the Passive RIS-aided Spectrum Sensing}
 The received signal at the ST for the passive RIS is given by
$\boldsymbol{\rm y}^{\rm pas}
=(\boldsymbol{{\rm h}}_{\rm d}+\boldsymbol{\rm H}\boldsymbol{\rm \Theta}\boldsymbol{\rm h}_{{\rm r}})\sqrt{p}s + \boldsymbol{\rm n}$,
where $p$ and $s$ denote the transmit power and the information signal of the PT, respectively.  $\boldsymbol{\rm \Theta}\triangleq{\textrm{diag}}([e^{j\theta_{1}},\cdots,e^{j\theta_{n}},\cdots e^{j\theta_{N}}]^H)$ is the diagonal phase-shift matrix, where $\theta_{n}\in[0,2\pi)$.  $\boldsymbol{\rm n}\sim\mathcal{CN}(0,\delta^2\boldsymbol{\rm I}_M)$ denotes the additive white Gaussian noise (AWGN) at the ST. The ST applies a receive beamformer $\boldsymbol{\rm w}$ with $\|\boldsymbol{\rm w}\|^2=1$ to equalize the received signal, which can be expressed as
$y^{\rm pas}=\boldsymbol{\rm w}^H\boldsymbol{\rm y}^{\rm pas}=\boldsymbol{\rm w}^H(\boldsymbol{{\rm h}}_{\rm d}+\boldsymbol{\rm H}\boldsymbol{\rm \Theta}\boldsymbol{\rm h}_{{\rm r}})\sqrt{p}s + \boldsymbol{\rm w}^H \boldsymbol{\rm n}$.

During the sensing interval, the ST can collect $I$ signal samples to determine whether the PT is active. Therefore, we can rewrite the received signal at the ST during $i$-th sampling instant as
\begin{equation}\label{eq3}
y^{\rm pas}(i)=\left\{
\begin{split}
\boldsymbol{\rm w}^H& \boldsymbol{\rm n}(i),~~~~~~~~~~~~~~~~~~~~~~~~~~~~~~~~~~~~\mathcal{H}_0\\
\boldsymbol{\rm w}^H&(\boldsymbol{{\rm h}}_{\rm d}+\boldsymbol{\rm H}\boldsymbol{\rm \Theta}\boldsymbol{\rm h}_{{\rm r}})\sqrt{p}s(i) +\boldsymbol{\rm w}^H \boldsymbol{\rm n}(i),~~~\mathcal{H}_1
\end{split}
\right.
\end{equation}
where $y^{\rm pas}(i)$ denotes the $i$-th sample, $\mathcal{H}_0$ denotes that the PT is inactive and $\mathcal{H}_1$ denotes that the PT is active. Energy detection is the most popular spectrum sensing scheme due to its easy implementation in practice. We apply the energy detector for spectrum detection in the proposed RIS-aided system. Let $f_s$ be the signal sampling frequency and $\tau$ be the spectrum sensing time. Thus, the number of samples $I$ can be given as $I=\tau f_s$. The test statistic for the energy detector is given by
$T(y)=\frac{1}{I}\sum_{i=1}^I|y(i)|^2$.
{\color{black}Under hypothesis $\mathcal{H}_0$, the test statistic $T(y)$ is a random variable whose probability density function (PDF)  is a Chi-square distribution with $2I$ degrees of freedom for the complex-valued case\cite{c1}.} Applying the central limit theorem (CLT), we have the following propositions.

\textit{Proposition 1:} When the number of samples is sufficiently large, for the passive RIS, the PDF of $T(y)$ under hypothesis $\mathcal{H}_0$ can be approximated by a Gaussian distribution\cite{c1} with the following mean $u_0^{\rm pas}=\delta^2$ and variance $v_0^{\rm pas}=\frac{\delta^4}{I}$.
%\begin{equation}\label{eq5}
%\begin{split}
%u_0^{\rm pas}&=\delta^2,~~
%%\end{split}
%%\end{equation}
%%\begin{equation}\label{eq61}
%%\begin{split}
%v_0^{\rm pas}=\frac{\delta^4}{I}.
%\end{split}
%\end{equation}
\begin{proof}
	Please refer to Appendix A.
\end{proof}

\textit{Proposition 2:} When the number of samples is sufficiently large, for the passive RIS, the PDF of $T(y)$ under hypothesis $\mathcal{H}_1$ can be approximated by a Gaussian distribution with the following mean $u_1^{\rm pas}$ and variance $v_1^{\rm pas}$
\begin{equation}\label{eq6}
\begin{split}
u_1^{\rm pas}&=(\gamma^{\rm act}+1)\delta^2=p|\boldsymbol{\rm w}^H(\boldsymbol{{\rm h}}_{\rm d}+\boldsymbol{\rm H}\boldsymbol{\rm \Theta}\boldsymbol{\rm h}_{{\rm r}})|^2+\delta^2,\\
%\end{split}
%\end{equation}
%\begin{equation}\label{eq61}
%\begin{split}
v_1^{\rm pas}&=\frac{1}{I}(p^2|\boldsymbol{\rm w}^H(\boldsymbol{{\rm h}}_{\rm d}+\boldsymbol{\rm H}\boldsymbol{\rm \Theta}\boldsymbol{\rm h}_{{\rm r}})|^4+\delta^4\\ &+2\delta^2p|\boldsymbol{\rm w}^H(\boldsymbol{{\rm h}}_{\rm d}+\boldsymbol{\rm H}\boldsymbol{\rm \Theta}\boldsymbol{\rm h}_{{\rm r}})|^2)=\frac{1}{I}(\gamma^{\rm pas}+1)^2\delta^4,
\end{split}
\end{equation}
where $\gamma^{\rm pas}=p|\boldsymbol{\rm w}^H(\boldsymbol{{\rm h}}_{\rm d}+\boldsymbol{\rm H}\boldsymbol{\rm \Theta}\boldsymbol{\rm h}_{{\rm r}})|^2/\delta^2$.
\begin{proof}
	Please refer to Appendix B.
\end{proof}

For an RIS-enhanced energy detection scheme, the decision rule at the ST can be given by $T(y)\underset{\mathcal{H}_0}{\overset{\mathcal{H}_1}{\gtrless}}\epsilon$,
where $\epsilon>0$  is the detection threshold. According to Propositions 1 and 2, the expressions for {\color{black}the detection probability $P_d^{\rm pas}$ and the probability of false alarm $P_f^{\rm pas}$ can be respectively given as}
\begin{equation}\label{eq9}
\begin{split}
P_f^{\rm pas}&=Q\left(\left(\frac{\epsilon}{\delta^2}-1\right)\sqrt{I}\right),\\
%\end{split}
%\end{equation}
%\begin{equation}\label{eq61}
%\begin{split}
P_d^{\rm pas}&=Q\left(\left(\frac{\epsilon}{\delta^2}-\gamma^{\rm pas}-1\right)\sqrt{\frac{I}{(1+\gamma^{\rm pas})^2}}\right),
\end{split}
\end{equation}
where $Q(x)=\frac{1}{\sqrt{2\pi}}\int_{x}^{\infty} \exp(-\frac{t^2}{2})\, dt$ is the complementary distribution function of the standard Gaussian distribution.
\begin{proof}
	Please refer to Appendix C.
\end{proof}

%\begin{figure}
%	\centering
%	\begin{subfigure}{0.49\linewidth}
%		\centering
%		\includegraphics[width=1.1\linewidth]{fig1.eps}
%		\caption{}
%		\label{fig1}%文中引用该图片代号
%	\end{subfigure}
%	\centering
%	\begin{subfigure}{0.49\linewidth}
%		\centering
%		\includegraphics[width=1.1\linewidth]{fig2.eps}
%		\caption{}
%		\label{fig2}%文中引用该图片代号
%	\end{subfigure}
%	\caption{(a) The detection probability versus the sensing SNR, $P_f^{\max}=0.1$. (b)  The probability of false alarm versus the sensing SNR, $P_d^{\min}=0.9$.}
%	\label{fig12}
%\end{figure}

\subsection{\color{black}System Model for the Active RIS-aided Spectrum Sensing}
Different from the passive RIS, the active RIS will amplify noise, which cannot be ignored. Therefore, we can rewrite the received signal at the ST during $i$-th sampling instant as
\begin{equation}\label{eq10}
y^{\rm act}(i)=\left\{
\begin{split}
\boldsymbol{\rm w}^H& \boldsymbol{\rm n}(i),~~~~~~~~~~~~~~~~~~~~~~~~~~~~~~~~~~\mathcal{H}_0\\
\boldsymbol{\rm w}^H&(\boldsymbol{{\rm h}}_{\rm d}+\boldsymbol{\rm H}\boldsymbol{\rm \Lambda}\boldsymbol{\rm \Theta}\boldsymbol{\rm h}_{{\rm r}})\sqrt{p}s(i) \\
&+ \boldsymbol{\rm w}^H\boldsymbol{\rm H}\boldsymbol{\rm \Lambda}\boldsymbol{\rm \Theta}\boldsymbol{\rm z}(i) +\boldsymbol{\rm w}^H \boldsymbol{\rm n}(i),~~~~~~\mathcal{H}_1
\end{split}
\right.
\end{equation}
where $\boldsymbol{\rm \Lambda}={\textrm{diag}}(\rho_{1},\cdots,\rho_{n},\cdots,\rho_{N})$ denotes the amplification factor matrix, where $\rho_{n}> 1$ denotes the amplification factor of the $n$-th reflecting element. $\boldsymbol{\rm z}\in\mathbb{C}^{N\times 1}$ {\color{black}is the noise introduced and amplified by the reflection-type amplifier\cite{i51},} which is assumed to follow the independent circularly symmetric complex Gaussian (CSCG) distribution, i.e., $\boldsymbol{\rm z}\sim\mathcal{CN}(\boldsymbol{0}, \sigma^2\boldsymbol{\rm I}_N)$. Applying the CLT, we have the following propositions.

\textit{Proposition 3:} When the number of samples is sufficiently large, for the active RIS, the PDF of $T(y)$ under hypothesis $\mathcal{H}_0$ can be approximated by a Gaussian distribution with the following mean $u_0^{\rm act}=\delta^2$ and variance $v_0^{\rm act}=\frac{\delta^4}{I}$.
%\begin{equation}\label{eq11}
%\begin{split}
%u_0^{\rm act}&=\delta^2,~~
%%\end{split}
%%\end{equation}
%%\begin{equation}\label{eq61}
%%\begin{split}
%v_0^{\rm act}=\frac{\delta^4}{I}.
%\end{split}
%\end{equation}
%We assume that the received signal $\boldsymbol{\rm w}^H\boldsymbol{\rm n}(i)$ is CSCG. In this case, $\mathbb{E}\{|\boldsymbol{\rm w}^H\boldsymbol{\rm n}(i)|^4\}=2\delta^4$. Substituting it into (\ref{eq11}), we can obtain $v_0^{\rm act}=\frac{\delta^4}{I}$.

\textit{Proposition 4:} When the number of samples is sufficiently large, for the active RIS, the PDF of $T(y)$ under hypothesis $\mathcal{H}_1$ can be approximated by a Gaussian distribution with the following mean $u_1^{\rm act}$ and variance $v_1^{\rm act}$
\begin{equation}\label{eq12}
\begin{split}
&u_1^{\rm act}=(\gamma^{\rm act}+1)(\sigma^2\|\boldsymbol{\rm w}^H\boldsymbol{{\rm H}}\boldsymbol{\rm \Lambda}\boldsymbol{\rm \Theta}\|^2+\delta^2)\\
&=p|\boldsymbol{\rm w}^H(\boldsymbol{{\rm h}}_{\rm d}+\boldsymbol{\rm H}\boldsymbol{\rm \Lambda}\boldsymbol{\rm \Theta}\boldsymbol{\rm h}_{{\rm r}})|^2+\sigma^2\|\boldsymbol{\rm w}^H\boldsymbol{{\rm H}}\boldsymbol{\rm \Lambda}\boldsymbol{\rm \Theta}\|^2+\delta^2,\\
%\end{split}
%\end{equation}
%\begin{equation}\label{eq61}
%\begin{split}
&v_1^{\rm act}{=}\frac{1}{I}(p^2|\boldsymbol{\rm w}^H(\boldsymbol{{\rm h}}_{\rm d}{+}\boldsymbol{\rm H}\boldsymbol{\rm \Lambda}\boldsymbol{\rm \Theta}\boldsymbol{\rm h}_{{\rm r}})|^4{+}\sigma^4\|\boldsymbol{\rm w}^H\boldsymbol{\rm H}\boldsymbol{\rm \Lambda}\boldsymbol{\rm \Theta}\|^4{+}\delta^4\\
&{+}2p\sigma^2|\boldsymbol{\rm w}^H(\boldsymbol{{\rm h}}_{\rm d}{+}\boldsymbol{\rm H}\boldsymbol{\rm \Lambda}\boldsymbol{\rm \Theta}\boldsymbol{\rm h}_{{\rm r}})|^2\|\boldsymbol{\rm w}^H\boldsymbol{\rm H}\boldsymbol{\rm \Lambda}\boldsymbol{\rm \Theta}\|^2\\
&{+}2\delta^2p|\boldsymbol{\rm w}^H(\boldsymbol{{\rm h}}_{\rm d}{+}\boldsymbol{\rm H}\boldsymbol{\rm \Lambda}\boldsymbol{\rm \Theta}\boldsymbol{\rm h}_{{\rm r}})|^2{+}2\delta^2\sigma^2\|\boldsymbol{\rm w}^H\boldsymbol{\rm H}\boldsymbol{\rm \Lambda}\boldsymbol{\rm \Theta}\|^2),
\end{split}
\end{equation}
where
$\gamma^{\rm act}=p|\boldsymbol{\rm w}^H(\boldsymbol{{\rm h}}_{\rm d}+\boldsymbol{\rm H}\boldsymbol{\rm \Lambda}\boldsymbol{\rm \Theta}\boldsymbol{\rm h}_{{\rm r}})|^2/(\sigma^2\|\boldsymbol{\rm w}^H\boldsymbol{{\rm H}}\boldsymbol{\rm \Lambda}\boldsymbol{\rm \Theta}\|^2+\delta^2)$. The proofs of propositions 3 and 4  are similar to Appendices A and B, and thus they are omitted for brevity. 
%Suppose that the received signal $\boldsymbol{\rm w}^H(\boldsymbol{{\rm h}}_{\rm d}+\boldsymbol{\rm H}\boldsymbol{\rm \Lambda}\boldsymbol{\rm \Theta}\boldsymbol{\rm h}_{{\rm r}})\sqrt{p}s(i)$ is CSCG, we have $\mathbb{E}\{|\boldsymbol{\rm w}^H(\boldsymbol{{\rm h}}_{\rm d}+\boldsymbol{\rm H}\boldsymbol{\rm \Lambda}\boldsymbol{\rm \Theta}\boldsymbol{\rm h}_{{\rm r}})\sqrt{p}s(i)|^4\}=2p^2|\boldsymbol{\rm w}^H(\boldsymbol{{\rm h}}_{\rm d}+\boldsymbol{\rm H}\boldsymbol{\rm \Lambda}\boldsymbol{\rm \Theta}\boldsymbol{\rm h}_{{\rm r}})|^4$, substituting it into (\ref{eq12}), we obtain $v_1^{\rm act}=\frac{1}{I}(p^2|\boldsymbol{\rm w}^H(\boldsymbol{{\rm h}}_{\rm d}+\boldsymbol{\rm H}\boldsymbol{\rm \Lambda}\boldsymbol{\rm \Theta}\boldsymbol{\rm h}_{{\rm r}})|^4+\sigma^4\|\boldsymbol{\rm w}^H\boldsymbol{\rm H}\boldsymbol{\rm \Lambda}\boldsymbol{\rm \Theta}\|^4+\delta^4+2p\sigma^2|\boldsymbol{\rm w}^H(\boldsymbol{{\rm h}}_{\rm d}+\boldsymbol{\rm H}\boldsymbol{\rm \Lambda}\boldsymbol{\rm \Theta}\boldsymbol{\rm h}_{{\rm r}})|^2\|\boldsymbol{\rm w}^H\boldsymbol{\rm H}\boldsymbol{\rm \Lambda}\boldsymbol{\rm \Theta}\|^2+2\delta^2p|\boldsymbol{\rm w}^H(\boldsymbol{{\rm h}}_{\rm d}+\boldsymbol{\rm H}\boldsymbol{\rm \Lambda}\boldsymbol{\rm \Theta}\boldsymbol{\rm h}_{{\rm r}})|^2+2\delta^2\sigma^2\|\boldsymbol{\rm w}^H\boldsymbol{\rm H}\boldsymbol{\rm \Lambda}\boldsymbol{\rm \Theta}\|^2)$. 
The expressions for the detection probability and the probability of false alarm can be respectively given as
\begin{equation}\label{eq14}
\begin{array}{l}
P_f^{\rm act}=Q\left(\left(\frac{\epsilon}{\delta^2}-1\right)\sqrt{I}\right),\\
%\end{array}
%\end{equation}
%\begin{equation}\label{eq61}
%\begin{array}
P_d^{\rm act}=Q((\frac{\epsilon}{p|\boldsymbol{\rm w}^H(\boldsymbol{{\rm h}}_{\rm d}+\boldsymbol{\rm H}\boldsymbol{\rm \Lambda}\boldsymbol{\rm \Theta}\boldsymbol{\rm h}_{{\rm r}})|^2{+}\sigma^2\|\boldsymbol{\rm w}^H\boldsymbol{{\rm H}}\boldsymbol{\rm \Lambda}\boldsymbol{\rm \Theta}\|^2{+}\delta^2}{-}1)\sqrt{I})\\
~~~~~~=Q((\frac{\epsilon}{\sigma^2\|\boldsymbol{\rm w}^H\boldsymbol{{\rm H}}\boldsymbol{\rm \Lambda}\boldsymbol{\rm \Theta}\|^2{+}\delta^2}{-}\gamma^{\rm act}{-}1)\frac{\sqrt{I}}{1+\gamma^{\rm act}}).
\end{array}
\end{equation}

\textbf{\textit{Remark 1:}} According to (\ref{eq9}) and (\ref{eq14}), we can see that the passive RIS can improve the sensing ability of the ST by constructing an additional reflecting link. Different from active RIS-aided wireless communication, the thermal noise in the active RIS-aided spectrum sensing will deteriorate wireless communication performance. However, the thermal noise in the active RIS-aided spectrum sensing systems can enhance the spectrum sensing.

{\color{black}The active RIS requires extra power to sustain its active load. Thus, it can allocate the remaining power to amplify the incident signal after covering the hardware power consumption for $N$ reflecting elements. However, it is essential to note that the amplification power is restricted by the total power budget available\cite{i5}.}
%It is worth noting that although the active RIS can amplify the received signal, it consumes more energy than the passive RIS. Thus, the amplification power of the active RIS is limited due to the total power budget at the active RIS. 
%Then, there exist four cases for the amplification power constraint, i.e., Case 1: When the PT is active and no false alarm is generated by the ST. Case 2: When the PT is inactive but it is detected by the ST. Case 3: When the PT is active but it is not detected by the ST. Case 4: When the PT is inactive and it is not detected  by the ST. 
The probabilistic amplification power constraint of the active RIS is
\begin{equation}\label{eq15}
\begin{split}
{\rm Pr}(\mathcal{H}_1)P_d^{\rm act}(p\|\boldsymbol{\rm \Lambda}\boldsymbol{\rm \Theta}\boldsymbol{\rm h}_{\rm r}\|^2+\sigma^2\|\boldsymbol{\rm \Lambda}\boldsymbol{\rm \Theta}\|_F^2)\leq P_{\rm RIS}^{\max},\\
%\end{split}
%\end{equation}
%\begin{equation}\label{eq61}
%\begin{split}
%&\textrm{Case 2:}~{\rm Pr}(\mathcal{H}_0)P_f^{\rm act}(p\|\boldsymbol{\rm \Lambda}\boldsymbol{\rm \Theta}\boldsymbol{\rm h}_{\rm r}\|^2+\sigma^2\|\boldsymbol{\rm \Lambda}\boldsymbol{\rm \Theta}\|_F^2)\leq P_{\rm RIS}^{\max},\\
%\end{split}
%\end{equation}
%\begin{equation}\label{eq61}
%\begin{split}
%&\textrm{Case 3:}~{\rm Pr}(\mathcal{H}_1)(1{-}P_d^{\rm act})(p\|\boldsymbol{\rm \Lambda}\boldsymbol{\rm \Theta}\boldsymbol{\rm h}_{\rm r}\|^2{+}\sigma^2\|\boldsymbol{\rm \Lambda}\boldsymbol{\rm \Theta}\|_F^2){\leq} P_{\rm RIS}^{\max},\\
%\end{split}
%\end{equation}
%\begin{equation}\label{eq61}
%\begin{split}
%&\textrm{Case 4:}~{\rm Pr}(\mathcal{H}_0)(1{-}P_f^{\rm act})(p\|\boldsymbol{\rm \Lambda}\boldsymbol{\rm \Theta}\boldsymbol{\rm h}_{\rm r}\|^2{+}\sigma^2\|\boldsymbol{\rm \Lambda}\boldsymbol{\rm \Theta}\|_F^2){\leq} P_{\rm RIS}^{\max},
\end{split}
\end{equation}
where ${\rm Pr}(\mathcal{H}_1)$ denotes the probability for which the PT is active. Correspondingly, ${\rm Pr}(\mathcal{H}_0)$ denotes the probability for which the PT is inactive. $P_{\rm RIS}^{\max}$ denotes the maximum power threshold of the active RIS.

\textbf{\textit{Remark 2:}} In fact, there exist four cases {\color{black}related to} the amplification power constraint, i.e., Case 1: When the PT is active and no false alarm is generated by the ST. Case 2: When the PT is {\color{black}inactive, it is} detected by the ST. Case 3: When the PT is {\color{black}active, it is} not detected by the ST. Case 4: When the PT is inactive, it is not detected by the ST. {\color{black}That is to say, there is no signal to be amplified for cases 2 and 4, and the amplification is applied for case 3. Thus, there is} only case 1 left for the amplification power constraint, i.e., (\ref{eq15}).

\section{\color{black}The Algorithm Design for the Passive RIS-aided Spectrum Sensing}
\subsection{Problem Formulation}
In this section, we focus on maximizing the detection probability by jointly optimizing the detection threshold, the receive beamforming vector, and the phase-shift matrix for passive RIS-aided spectrum sensing. By considering the maximum probability constraint of false alarm and the phase-shift constraint,  the detection probability maximization problem is expressed as follows
\begin{equation}\label{eq16}
\begin{split}
\max\limits_{\mbox{\scriptsize$\begin{array}{c} 
		\boldsymbol{\rm \Theta},\boldsymbol{\rm w},\epsilon
		\end{array}$}} 
&P_{d}^{\rm pas}\\
s.t.~{C_1}: &P_f^{\rm pas}\leq P_f^{\max},~~{C_2}:0\leq\theta_{n}<2\pi,\\
{C_3}:&\|\boldsymbol{{\rm w}}\|^2=1,
\end{split}
\end{equation}
where $P_f^{\max}$ denotes the maximum probability of false alarm. Specifically, $C_1$ states that the maximum probability constraint of false alarm, $C_2$ is the unit-modulus constraint, {\color{black}and $C_3$ denotes} the receive beamforming constraint.

\textbf{\textit{Remark 3:}} {\color{black}It is worth noting that although passive RIS-aided spectrum sensing has been considered in \cite{e1,e2,e3,e4,e5}}, these works focus more on the performance analysis of detection probability or the throughput of the secondary network, which are completely different from this work where we aim at enhancing the sensing performance. Thus, the methods developed in \cite{e1,e2,e3,e4,e5} are not applicable here.

\subsection{Problem Transformation}
We note that problem (\ref{eq16}) is a non-convex optimization problem due to the Q function and coupled variables. In the following, we develop a sub-optimal AO-based iterative algorithm to solve problem (\ref{eq16}). For the given $\boldsymbol{{\rm w}}$ and $\boldsymbol{\rm \Theta}$, the receive beamforming vector can be shown as
\begin{equation}\label{eq17}
\begin{split}
\boldsymbol{\rm w}^*=\frac{\boldsymbol{{\rm h}}_{\rm d}+\boldsymbol{\rm H}\boldsymbol{\rm \Theta}\boldsymbol{\rm h}_{{\rm r}}}{\|\boldsymbol{{\rm h}}_{\rm d}+\boldsymbol{\rm H}\boldsymbol{\rm \Theta}\boldsymbol{\rm h}_{{\rm r}}\|}.
\end{split}
\end{equation}
With the beamforming vector, the SNR expression can be rewritten as
%\begin{equation}\label{eq18}
%\begin{split}
$\bar \gamma^{\rm pas}{=}(p\|\boldsymbol{{\rm h}}_{\rm d}+\boldsymbol{\rm H}\boldsymbol{\rm \Theta}\boldsymbol{\rm h}_{{\rm r}}\|^2)/\delta^2$.
%\end{split}
%\end{equation}
It is worth noting that the Q function is a monotonically non-increasing function, and the detection probability can be maximized only when $\epsilon$ takes the minimum value. For given $\boldsymbol{{\rm w}}$ and $\boldsymbol{\rm \Theta}$, based on the monotonicity of the Q function and $C_1$, we know that $\epsilon$ must satisfy the condition
$\epsilon\geq \frac{\delta^2 Q^{-1}(P_f^{\max})}{\sqrt{I}}+\delta^2$.
Thus, we can design  the detection threshold as
\begin{equation}\label{eq20}
\begin{split}
\epsilon^* = \frac{\delta^2 Q^{-1}(P_f^{\max})}{\sqrt{I}}+\delta^2.
\end{split}
\end{equation}

Let's substitute $\boldsymbol{{\rm w}}^*$ and $\epsilon^*$ into problem (\ref{eq16}). Based on the monotonicity of the Q function, we can transform problem (\ref{eq16}) into the following problem
\begin{equation}\label{eq22}
\begin{split}
\max\limits_{\mbox{\scriptsize$\begin{array}{c} 
		\boldsymbol{\rm \Theta}
		\end{array}$}} 
~~~&\bar\gamma^{\rm pas}\\
s.t.~&{C_2}.
\end{split}
\end{equation}
Thus, we can obtain the optimal phase shift\cite{i1}, i.e.,
\begin{equation}\label{eq23}
\begin{split}
\boldsymbol{{\rm \theta}}^*=e^{j\{{\rm arg}(\boldsymbol{\rm w}^H\boldsymbol{{\rm h}}_{\rm d})-{\rm arg}(\boldsymbol{\rm w}^H\boldsymbol{{\rm H}}{\rm diag}(\boldsymbol{{\rm h}}_{\rm r}))\}}.
\end{split}
\end{equation}
Based on the above solutions, we can obtain the maximum detection probability for the passive RIS.

\section{\color{black}The Algorithm Design for the Active RIS-aided Spectrum Sensing}
The traditional passive RIS suffers from the “double fading” effect, which has become a major bottleneck in restricting the performance of passive RIS-aided spectrum sensing. Different from the passive RIS, the optimization process for the active RIS is more complex since it involves optimizing not only the phase shift but also the amplification factor. Furthermore, the active RIS has higher power consumption and is subject to the maximum power constraint of the active RIS. This means that the spectrum sensing algorithm for the passive RIS cannot be directly applied to the active {\color{black}RIS. Thus, new spectrum sensing methods need} to be designed.
\subsection{Problem Formulation}
Similar to the passive RIS, we also investigate the detection probability maximization problem for the active RIS.  Specifically,  the detection probability maximization problem is
\begin{equation}\label{eq24}
\begin{split}
&\max\limits_{\mbox{\scriptsize$\begin{array}{c} 
		\boldsymbol{\rm \Theta},\boldsymbol{\rm w},\rho_{n},\epsilon
		\end{array}$}} 
P_d^{\rm act}\\
\!\!\!\!s.t.~{C'_1}: &P_f^{\rm act}\leq P_f^{\max},\\
{C'_2}:&{\rm Pr}(\mathcal{H}_1)P_d^{\rm act}(p\|\boldsymbol{\rm \Lambda}\boldsymbol{\rm \Theta}\boldsymbol{\rm h}_{\rm r}\|^2{+}\sigma^2\|\boldsymbol{\rm \Lambda}\boldsymbol{\rm \Theta}\|_F^2){\leq} P_{\rm RIS}^{\max},\\
{C'_3}:&\|\boldsymbol{{\rm w}}\|^2=1,
\end{split}
\end{equation}
where $C'_2$ guarantees the maximum power constraint of the active RIS. 

\textbf{\textit{Remark 4:}} To the best of our knowledge, this is the first work to improve the sensing performance for active RIS-aided spectrum sensing while ensuring a target probability of false alarm and probabilistic power consumption for the active RIS.
%It is noted that, although active RIS-aided spectrum sensing was considered in \cite{f1}, which is completely different from this work regarding the problem formulation and resulting solution methodology. In fact, \cite{f1} investigated the combination of spectrum sensing and wireless communication, and its focus is on the trade-off rather than the sensing performance, which is also in contrast of our works in which focus on improving the sensing performance while ensuring a target probability of false alarm and probabilistic power consumption for the active RIS. Thus, the methods developed in \cite{f1} are also not applicable here.
Compared to the passive RIS, although the active RIS avoids the unit-modulus constraint, it also introduces the additional probabilistic non-convex constraint $C'_2$, which leads to the coupling between the optimization variables, i.e., the amplification factor and the phase-shift matrix. 
%The Q function is neither convex nor concave, unlike the passive RIS, we can take advantage of the monotonicity of the Q function. However, $C_2$ introduced by the active RIS makes it difficult to take advantage of its monotonicity. 
Furthermore, the product between the Q function and other optimization variables further aggravates the coupling among variables. In the following, we propose two algorithms to solve problem (\ref{eq24}), i.e., one-stage beamforming optimization algorithm and two-stage beamforming optimization algorithm.

\subsection{One-stage Beamforming Optimization}
First, we consider optimizing the detection threshold. Similar to the passive RIS, the probability of false alarm can be limited and guaranteed by designing the detection threshold. Thus, the probability of false alarm can be guaranteed when the detection threshold is designed as follows
%\begin{equation}\label{eq25}
%\begin{split}
%\epsilon\geq \frac{\delta^2 Q^{-1}(P_f^{\max})}{\sqrt{I}}+\delta^2.
%\end{split}
%\end{equation}
%Then, we can design  the detection threshold as
\begin{equation}\label{eq26}
\begin{split}
\epsilon^* = \frac{\delta^2 Q^{-1}(P_f^{\max})}{\sqrt{I}}+\delta^2,
\end{split}
\end{equation}
where the detection threshold of the active RIS is the same as that of the passive RIS. Next, let us observe the optimization problem (\ref{eq24}). The main difficulty in (\ref{eq24}) is the Q function in $C'_2$. Thus,  one feasible solution to deal with it is to relax $C'_2$. Since $0\leq P_d^{\rm act}\leq 1$, 
%we have the following inequality hold
%\begin{equation}\label{eq27}
%\begin{split}
%{\rm Pr}(\mathcal{H}_1)P_d^{\rm act}(p\|\boldsymbol{\rm \Lambda}\boldsymbol{\rm \Theta}\boldsymbol{\rm h}_{\rm r}\|^2+\sigma^2\|\boldsymbol{\rm \Lambda}\boldsymbol{\rm \Theta}\|_F^2)\\
%\leq {\rm Pr}(\mathcal{H}_1)(p\|\boldsymbol{\rm \Lambda}\boldsymbol{\rm \Theta}\boldsymbol{\rm h}_{\rm r}\|^2+\sigma^2\|\boldsymbol{\rm \Lambda}\boldsymbol{\rm \Theta}\|_F^2).
%\end{split}
%\end{equation}
{\color{black}we can transform $C'_2$ into}
${\bar C'_2}:{\rm Pr}(\mathcal{H}_1)(p\|\boldsymbol{\rm \Lambda}\boldsymbol{\rm \Theta}\boldsymbol{\rm h}_{\rm r}\|^2+\sigma^2\|\boldsymbol{\rm \Lambda}\boldsymbol{\rm \Theta}\|_F^2)\leq P_{\rm RIS}^{\max}$.
{\color{black}It is worth noting that} the original constraint $C'_2$ can be guaranteed when $\bar C'_2$ holds. Since the Q function is a monotonically non-increasing function, the problem (\ref{eq24}) can be reformulated as
\begin{equation}\label{eq29}
\begin{split}
\!\!\max\limits_{\mbox{\scriptsize$\begin{array}{c} 
		\boldsymbol{\rm \Theta},\boldsymbol{\rm w},\rho_{n}
		\end{array}$}} 
&p|\boldsymbol{\rm w}^H(\boldsymbol{{\rm h}}_{\rm d}{+}\boldsymbol{\rm H}\boldsymbol{\rm \Lambda}\boldsymbol{\rm \Theta}\boldsymbol{\rm h}_{{\rm r}})|^2{+}\sigma^2\|\boldsymbol{\rm w}^H\boldsymbol{{\rm H}}\boldsymbol{\rm \Lambda}\boldsymbol{\rm \Theta}\|^2{+}\delta^2\\
s.t.~&{\bar C'_2},C'_3.
\end{split}
\end{equation}
Compared to the problem (\ref{eq24}), the problem (\ref{eq29}) has been transformed into a more tractable form. However, the problem (\ref{eq29}) remains an intractable one because the receive beamforming vector is coupled not only with the amplification {\color{black}factor but} also with the phase shift. 
%Although an alternate optimization method is used to decouple each variable so that multiple variables can be optimized separately. However, the objective function after using the alternate optimization method is a convex function, the problem is still a non-convex problem. Thus, we need to find other feasible methods. 
{\color{black}It is worth noting that} $\boldsymbol{\rm \Lambda}$ and $\boldsymbol{\rm \Theta}$ are always coupled together in the product form. Thus, we define a new variable $\boldsymbol{\rm \Phi}=\boldsymbol{\rm \Lambda}\boldsymbol{\rm \Theta}$. Then, the objective function can be rewritten as
%\begin{equation}\label{eq30}
%\begin{split}
$p|\boldsymbol{\rm w}^H(\boldsymbol{{\rm h}}_{\rm d}+\boldsymbol{\rm H}\boldsymbol{\rm \Phi}\boldsymbol{\rm h}_{{\rm r}})|^2+\sigma^2\|\boldsymbol{\rm w}^H\boldsymbol{{\rm H}}\boldsymbol{\rm \Phi}\|^2+\delta^2$.
%\end{split}
%\end{equation}
In order to make the problem tractable, we introduce an additional variable $\boldsymbol{\rm u}=[u_1,\cdots,u_n,\cdots,u_N]^H$, where $u_n=\rho_{n}e^{j\theta_n}$. The first term of the objective function can be transformed into
$p|\boldsymbol{\rm w}^H(\boldsymbol{{\rm h}}_{\rm d}+\boldsymbol{\rm H}\boldsymbol{\rm \Phi}\boldsymbol{\rm h}_{{\rm r}})|^2=p{\rm Tr}(\boldsymbol{\rm W}\boldsymbol{\rm H}_{\rm dr}\boldsymbol{\rm O}\boldsymbol{\rm H}_{\rm dr}^H)$,
where $\boldsymbol{\rm W}=\boldsymbol{\rm w}\boldsymbol{\rm w}^H$, $\boldsymbol{\rm H}_{\rm dr}=[\boldsymbol{\rm H}{\rm diag}(\boldsymbol{\rm h}_{{\rm r}}), \boldsymbol{{\rm h}}_{\rm d}]$, $\boldsymbol{\rm O}=\boldsymbol{\rm o}\boldsymbol{\rm o}^H$ where $\boldsymbol{\rm o}^H=[\boldsymbol{\rm u}^H,1]$. It can be observed that there still exists the product of matrices in $p{\rm Tr}(\boldsymbol{\rm W}\boldsymbol{\rm H}_{\rm dr}\boldsymbol{\rm O}\boldsymbol{\rm H}_{\rm dr}^H)$. To this end, we can rewrite the term ${\rm Tr}(\boldsymbol{\rm W}\boldsymbol{\rm H}_{\rm dr}\boldsymbol{\rm O}\boldsymbol{\rm H}_{\rm dr}^H)$ by its equivalent form of the difference of convex (DC), which facilitates the application of the inner approximation. In particular, the term ${\rm Tr}(\boldsymbol{\rm W}\boldsymbol{\rm H}_{\rm dr}\boldsymbol{\rm O}\boldsymbol{\rm H}_{\rm dr}^H)$ can be rewritten as\cite{i2}:
${\rm Tr}(\boldsymbol{\rm W}\boldsymbol{\rm H}_{\rm dr}\boldsymbol{\rm O}\boldsymbol{\rm H}_{\rm dr}^H)=\frac{1}{2}\|\boldsymbol{\rm W}+\boldsymbol{\rm H}_{\rm dr}\boldsymbol{\rm O}\boldsymbol{\rm H}_{\rm dr}^H\|_F^2-\frac{1}{2}\|\boldsymbol{\rm W}\|_F^2-\frac{1}{2}\|\boldsymbol{\rm H}_{\rm dr}\boldsymbol{\rm O}\boldsymbol{\rm H}_{\rm dr}^H\|_F^2$,
where $-\frac{1}{2}\|\boldsymbol{\rm W}\|_F^2$ and $-\frac{1}{2}\|\boldsymbol{\rm H}_{\rm dr}\boldsymbol{\rm O}\boldsymbol{\rm H}_{\rm dr}^H\|_F^2$ are concave function with respect to $\boldsymbol{\rm W}$ and $\boldsymbol{\rm O}$, respectively, while $\frac{1}{2}\|\boldsymbol{\rm W}+\boldsymbol{\rm H}_{\rm dr}\boldsymbol{\rm O}\boldsymbol{\rm H}_{\rm dr}^H\|_F^2$ is convex with respect to $\boldsymbol{\rm W}$ and $\boldsymbol{\rm O}$. Thus, we have
\begin{equation}\label{eq33}
\begin{split}
&\|\boldsymbol{\rm W}{+}\boldsymbol{\rm H}_{\rm dr}\boldsymbol{\rm O}\boldsymbol{\rm H}_{\rm dr}^H\|_F^2{\geq} \|\boldsymbol{\rm W}^{(l)}{+}\boldsymbol{\rm H}_{\rm dr}\boldsymbol{\rm O}^{(l)}\boldsymbol{\rm H}_{\rm dr}^H\|_F^2\\
&{+}{\rm Tr}\left((2\boldsymbol{\rm W}^{(l)}{+}2\boldsymbol{\rm H}_{\rm dr}\boldsymbol{\rm O}^{(l)}\boldsymbol{\rm H}_{\rm dr}^H)^H(\boldsymbol{\rm W}{-}\boldsymbol{\rm W}^{(l)})\right)\\
&{+}{\rm Tr}\left( ( 2\boldsymbol{\rm H}_{\rm dr}^H\boldsymbol{\rm W}^{(l)}\boldsymbol{\rm H}_{\rm dr}  {+}2\boldsymbol{\rm H}_{\rm dr}^H\boldsymbol{\rm H}_{\rm dr}\boldsymbol{\rm O}^{(l)}\boldsymbol{\rm H}_{\rm dr}^H\boldsymbol{\rm H}_{\rm dr}    )^H(\boldsymbol{\rm O}{-}\boldsymbol{\rm O}^{(l)})         \right),
\end{split}
\end{equation}
where $\boldsymbol{\rm W}^{(l)}$ and $\boldsymbol{\rm O}^{(l)}$ are the solutions obtained in the $l$-th iteration. Then, the first term of the objective function has been transformed into a concave function. In the following, we will deal with the second term of the objective function. Then, we have
$\sigma^2\|\boldsymbol{\rm w}^H\boldsymbol{{\rm H}}\boldsymbol{\rm \Phi}\|^2=\sigma^2{\rm Tr}(\boldsymbol{\rm \Phi}\boldsymbol{\rm \Phi}^H\boldsymbol{\rm H}^H\boldsymbol{\rm W}\boldsymbol{\rm H})$.
It can be observed that $\boldsymbol{\rm W}$ and $\boldsymbol{\rm \Phi}$ are coupled with each other. Then, we introduce a new variable $\boldsymbol{\rm \Psi}=\boldsymbol{\rm \Phi}\boldsymbol{\rm \Phi}^H$, thus ${\rm Tr}(\boldsymbol{\rm \Phi}\boldsymbol{\rm \Phi}^H\boldsymbol{\rm H}^H\boldsymbol{\rm W}\boldsymbol{\rm H})$ can be transformed into ${\rm Tr}(\boldsymbol{\rm W}\boldsymbol{\rm H}\boldsymbol{\rm \Psi}\boldsymbol{\rm H}^H)$. Furthermore, based on the same method above, we can rewrite ${\rm Tr}(\boldsymbol{\rm W}\boldsymbol{\rm H}\boldsymbol{\rm \Psi}\boldsymbol{\rm H}^H)$ as the form of DC. In particular, we have
${\rm Tr}(\boldsymbol{\rm W}\boldsymbol{\rm H}\boldsymbol{\rm \Psi}\boldsymbol{\rm H}^H)=\frac{1}{2}\|\boldsymbol{\rm W}+\boldsymbol{\rm H}\boldsymbol{\rm \Psi}\boldsymbol{\rm H}^H\|_F^2-\frac{1}{2}\|\boldsymbol{\rm W}\|_F^2-\frac{1}{2}\|\boldsymbol{\rm H}\boldsymbol{\rm \Psi}\boldsymbol{\rm H}^H\|_F^2$.
Similarly, we also need to obtain the concave form of the  term $\|\boldsymbol{\rm W}+\boldsymbol{\rm H}\boldsymbol{\rm \Psi}\boldsymbol{\rm H}^H\|_F^2$, which can be given by
\begin{equation}\label{eq36}
\begin{split}
&\|\boldsymbol{\rm W}+\boldsymbol{\rm H}\boldsymbol{\rm \Psi}\boldsymbol{\rm H}^H\|_F^2\geq \|\boldsymbol{\rm W}^{(l)}+\boldsymbol{\rm H}\boldsymbol{\rm \Psi}^{(l)}\boldsymbol{\rm H}^H\|_F^2\\
&+{\rm Tr}\left((2\boldsymbol{\rm W}^{(l)}+2\boldsymbol{\rm H}\boldsymbol{\rm \Psi}^{(l)}\boldsymbol{\rm H}^H)^H(\boldsymbol{\rm W}-\boldsymbol{\rm W}^{(l)})\right)\\
&+{\rm Tr}\left( ( 2\boldsymbol{\rm H}^H\boldsymbol{\rm W}^{(l)}\boldsymbol{\rm H}  {+}2\boldsymbol{\rm H}^H\boldsymbol{\rm H}\boldsymbol{\rm \Psi}^{(l)}\boldsymbol{\rm H}^H\boldsymbol{\rm H}    )^H(\boldsymbol{\rm \Psi}{-}\boldsymbol{\rm \Psi}^{(l)})         \right),
\end{split}
\end{equation}
where $\boldsymbol{\rm \Psi}^{(l)}$ is the solution obtained in the $l$-th iteration.
Then, the objective function is transformed into a concave form, i.e., (\ref{eq37}), which is shown at the top of this page. 
\begin{figure*}\vspace{-30pt}
	\begin{equation}\label{eq37}
	\begin{array}{l}
	p\left\{\frac{1}{2}\|\boldsymbol{\rm W}^{(l)}+\boldsymbol{\rm H}_{\rm dr}\boldsymbol{\rm O}^{(l)}\boldsymbol{\rm H}_{\rm dr}^H\|_F^2+{\rm Tr}\left((\boldsymbol{\rm W}^{(l)}+\boldsymbol{\rm H}_{\rm dr}\boldsymbol{\rm O}^{(l)}\boldsymbol{\rm H}_{\rm dr}^H)^H(\boldsymbol{\rm W}-\boldsymbol{\rm W}^{(l)})\right)-\frac{1}{2}\|\boldsymbol{\rm W}\|_F^2\right.\\
	\left.-\frac{1}{2}\|\boldsymbol{\rm H}_{\rm dr}\boldsymbol{\rm O}\boldsymbol{\rm H}_{\rm dr}^H\|_F^2+{\rm Tr}\left( ( \boldsymbol{\rm H}_{\rm dr}^H\boldsymbol{\rm W}^{(l)}\boldsymbol{\rm H}_{\rm dr}  +\boldsymbol{\rm H}_{\rm dr}^H\boldsymbol{\rm H}_{\rm dr}\boldsymbol{\rm O}^{(l)}\boldsymbol{\rm H}_{\rm dr}^H\boldsymbol{\rm H}_{\rm dr}    )^H(\boldsymbol{\rm O}-\boldsymbol{\rm O}^{(l)})         \right)    \right\}\ \\
	{+}  \sigma^2\left\{\frac{1}{2}\|\boldsymbol{\rm W}^{(l)}{+}\boldsymbol{\rm H}\boldsymbol{\rm \Psi}^{(l)}\boldsymbol{\rm H}^H\|_F^2{+}{\rm Tr}\left((\boldsymbol{\rm W}^{(l)}{+}\boldsymbol{\rm H}\boldsymbol{\rm \Psi}^{(l)}\boldsymbol{\rm H}^H)^H(\boldsymbol{\rm W}{-}\boldsymbol{\rm W}^{(l)})\right){+}{\rm Tr}( ( \boldsymbol{\rm H}^H\boldsymbol{\rm W}^{(l)}\boldsymbol{\rm H} \right. \\
	\left.+\boldsymbol{\rm H}^H\boldsymbol{\rm H}\boldsymbol{\rm \Psi}^{(l)}\boldsymbol{\rm H}^H\boldsymbol{\rm H}    )^H(\boldsymbol{\rm \Psi}-\boldsymbol{\rm \Psi}^{(l)})         ) -\frac{1}{2}\|\boldsymbol{\rm W}\|_F^2-\frac{1}{2}\|\boldsymbol{\rm H}\boldsymbol{\rm \Psi}\boldsymbol{\rm H}^H\|_F^2  \right\} +\delta^2\triangleq\mathcal{F}(\boldsymbol{\rm W},\boldsymbol{\rm O},\boldsymbol{\rm \Psi}).
	\end{array}
	\end{equation}
		\hrulefill
\end{figure*}
However, although the objective function has been converted into a tractable form, the additional equation constraints are also introduced, which are non-convex, i.e., $\boldsymbol{\rm W}=\boldsymbol{\rm w}\boldsymbol{\rm w}^H$, $\boldsymbol{\rm \Psi}=\boldsymbol{\rm \Phi}\boldsymbol{\rm \Phi}^H$, and $\boldsymbol{\rm O}=\boldsymbol{\rm o}\boldsymbol{\rm o}^H$. The three equalities can be equivalently written as the following constraints, respectively\cite{i3}
\begin{equation}\label{eq38}
\begin{split}
&C'_{4a}:\begin{bmatrix}
\boldsymbol{\rm W}& \boldsymbol{\rm w}\\  
\boldsymbol{\rm w}^H & 1\\ 
\end{bmatrix}
{\succeq} \boldsymbol{0}, C'_{4b}: {\rm Tr}(\boldsymbol{\rm W}{-}\boldsymbol{\rm w}\boldsymbol{\rm w}^H){\leq} 0, \\    
%\end{equation}
%\begin{equation}\label{eq24}
&C'_{5a}:\begin{bmatrix}
\boldsymbol{\rm \Psi}& \boldsymbol{\rm \Phi}\\  
\boldsymbol{\rm \Phi}^H & 1\\ 
\end{bmatrix}
{\succeq} \boldsymbol{0},C'_{5b}: {\rm Tr}(\boldsymbol{\rm \Psi}{-}\boldsymbol{\rm \Phi}\boldsymbol{\rm \Phi}^H){\leq} 0,\\      
%\end{equation}
%\begin{equation}\label{eq24}
&C'_{6a}:\begin{bmatrix}
\boldsymbol{\rm O}& \boldsymbol{\rm o}\\  
\boldsymbol{\rm o}^H & 1\\ 
\end{bmatrix} \succeq \boldsymbol{0},~~ C'_{6b}: {\rm Tr}(\boldsymbol{\rm O}-\boldsymbol{\rm o}\boldsymbol{\rm o}^H)\leq 0.  
\end{split} 
\end{equation}
We note that $C'_{4b}$, $C'_{5b}$, and $C'_{6b}$ are still non-convex. To deal with this obstacle, based on the first-order Taylor approximation, the lower bounds of ${\rm Tr}(\boldsymbol{\rm w}\boldsymbol{\rm w}^H)$, ${\rm Tr}(\boldsymbol{\rm \Phi}\boldsymbol{\rm \Phi}^H)$, and ${\rm Tr}(\boldsymbol{\rm o}\boldsymbol{\rm o}^H)$ can be derived as ${\rm Tr}(\boldsymbol{\rm w}\boldsymbol{\rm w}^H)\geq -\|\boldsymbol{\rm w}^{(l)}\|^2+2{\rm Tr}((\boldsymbol{\rm w}^{(l)})^H\boldsymbol{\rm w})$,
%\end{split}
%\end{equation}
%\begin{equation}\label{eq25}
%\begin{split}
${\rm Tr}(\boldsymbol{\rm \Phi}\boldsymbol{\rm \Phi}^H)\geq -\|\boldsymbol{\rm \Phi}^{(l)}\|^2+2{\rm Tr}((\boldsymbol{\rm \Phi}^{(l)})^H\boldsymbol{\rm \Phi})$,
%\end{split}
%\end{equation}
%\begin{equation}\label{eq25}
%\begin{split}
and ${\rm Tr}(\boldsymbol{\rm o}\boldsymbol{\rm o}^H)\geq -\|\boldsymbol{\rm o}^{(l)}\|^2+2{\rm Tr}((\boldsymbol{\rm o}^{(l)})^H\boldsymbol{\rm o})$, respectively,
where $\boldsymbol{\rm o}^{(l)}$, $\boldsymbol{\rm w}^{(l)}$, and $\boldsymbol{\rm \Phi}^{(l)}$ are the solutions obtained in the $l$-th iteration. Thus, based on the lower bounds above, we have
\begin{equation}\label{eq40}
\begin{split}
&\bar C'_{4b}:{\rm Tr}(\boldsymbol{\rm W})\leq -\|\boldsymbol{\rm w}^{(l)}\|^2+2{\rm Tr}((\boldsymbol{\rm w}^{(l)})^H\boldsymbol{\rm w}),\\
%\end{split}
%\end{equation}
%\begin{equation}\label{eq26}
%\begin{split}
&\bar C'_{5b}:{\rm Tr}(\boldsymbol{\rm \Psi})\leq -\|\boldsymbol{\rm \Phi}^{(l)}\|^2+2{\rm Tr}((\boldsymbol{\rm \Phi}^{(l)})^H\boldsymbol{\rm \Phi}),\\
%\end{split}
%\end{equation}
%\begin{equation}\label{eq26}
%\begin{split}
&\bar C'_{6b}:{\rm Tr}(\boldsymbol{\rm O})\leq -\|\boldsymbol{\rm o}^{(l)}\|^2+2{\rm Tr}((\boldsymbol{\rm o}^{(l)})^H\boldsymbol{\rm o}).
\end{split}
\end{equation}
Furthermore, based on the introduced variable, $\bar C'_2$ can be rewritten as the following convex constraint
${\hat C'_2}:{\rm Pr}(\mathcal{H}_1)(p\|\boldsymbol{\rm \Phi}\boldsymbol{\rm h}_{\rm r}\|^2+\sigma^2\|\boldsymbol{\rm \Phi}\|_F^2)\leq P_{\rm RIS}^{\max}$.
Then, problem (\ref{eq29}) can be transformed into the following convex problem
\begin{equation}\label{eq42}
\begin{split}
&\max\limits_{\mbox{\scriptsize$\begin{array}{c} 
		\boldsymbol{\rm w},	\boldsymbol{\rm W},\boldsymbol{\rm o},\boldsymbol{\rm O},\boldsymbol{\rm \Phi},\boldsymbol{\rm \Psi}
		\end{array}$}} 
~\mathcal{F}(\boldsymbol{\rm W},\boldsymbol{\rm O},\boldsymbol{\rm \Psi})\\
s.t.&\bar C'_3:{\rm Tr}(\boldsymbol{{\rm W}}){=}1,\boldsymbol{{\rm W}}{\succeq} \boldsymbol{0},\\
&C'_7:\boldsymbol{\rm o}_n{=}\boldsymbol{\rm \Phi}_{n,n},\boldsymbol{\rm O}{\succeq} \boldsymbol{0},\boldsymbol{\rm \Psi}\succeq \boldsymbol{0},\\
&\hat C'_2,C'_{4a}, \bar C'_{4b},C'_{5a}, \bar C'_{5b},C'_{6a}, \bar C'_{6b}.
%\boldsymbol{\rm O}\succeq \boldsymbol{0},\boldsymbol{\rm \Psi}\succeq \boldsymbol{0},
\end{split}
\end{equation}

\subsection{Two-stage Beamforming Optimization}
Let us recall the one-stage beamforming optimization algorithm, which scales the constraint $C'_2$. Although the one-stage beamforming optimization algorithm sidesteps the coupling between the Q function and optimization variables by scaling, it may squeeze the feasible region and may not lead to a satisfactory sensing performance. 
To facilitate performance improvement,  in the following,  we propose a two-stage optimization algorithm. We first transform problem (\ref{eq24}) into the following problem
\begin{equation}\label{eq43}
\begin{split}
&\max\limits_{\mbox{\scriptsize$\begin{array}{c} 
		\boldsymbol{\rm \Theta},\boldsymbol{\rm w},\rho_{n},\epsilon,t
		\end{array}$}} 
t\\
s.t.{C'_1}: &P_f^{\rm act}{\leq} P_f^{\max},\\
{C'_2}:&{\rm Pr}(\mathcal{H}_1)t(p\|\boldsymbol{\rm \Lambda}\boldsymbol{\rm \Theta}\boldsymbol{\rm h}_{\rm r}\|^2{+}\sigma^2\|\boldsymbol{\rm \Lambda}\boldsymbol{\rm \Theta}\|_F^2){\leq} P_{\rm RIS}^{\max},\\
{C'_3}:&\|\boldsymbol{{\rm w}}\|^2{=}1,~~~~C'_4:P_d^{\rm act}{\geq} t,
\end{split}
\end{equation}
where $t$ is an auxiliary variable. Note that (\ref{eq43}) is still non-convex since the inequality constraint set is not convex in $\{\boldsymbol{\rm \Theta},\boldsymbol{\rm w},\rho_{n},\epsilon,t\}$.  In particular, {\color{black}the two-stage beamforming optimization algorithm consists of the outer and inner loops.}  For any fixed $t$ in the outer loop, we can optimize other variables $\boldsymbol{\rm \Theta},\boldsymbol{\rm w},\rho_{n},\epsilon$ in the inner-loop optimization, which can effectively decouple the Q function. Given the variables $\boldsymbol{\rm \Theta},\boldsymbol{\rm w},\rho_{n},\epsilon$, we can optimize the auxiliary variable $t$ in the outer loop, which promotes us to apply the bisection method to update $t$ when we find the upper bound and the lower bound on $t$.

\subsubsection{Inner-loop Optimization}
For given $t$, problem (\ref{eq43}) becomes a feasibility problem. In particular, the feasibility problem can be formulated as follows
\begin{equation}\label{eq44}
\begin{split}
\textrm{\textbf{Find}} ~~~&\{\boldsymbol{\rm \Theta},\boldsymbol{\rm w},\rho_{n},\epsilon\}\\
s.t.~&{C'_1}-C'_4.
\end{split}
\end{equation}
Note that the convex constraint set can be denoted by
$\mathcal{S}=\{\{\boldsymbol{\rm \Theta},\boldsymbol{\rm w},\rho_{n},\epsilon\}|({C'_1}-C'_4)\}$.
Similarly, we can design the detection threshold as $\epsilon^* = \frac{\delta^2 Q^{-1}(P_f^{\max})}{\sqrt{I}}+\delta^2$ to ensure the probability of false alarm.
We substitute the detection threshold $\epsilon^*$ into $C'_4$, then $C'_4$ can be transformed into
\begin{equation}\label{eq47}
\begin{split}
&\bar C'_4:(p|\boldsymbol{\rm w}^H(\boldsymbol{{\rm h}}_{\rm d}{+}\boldsymbol{\rm H}\boldsymbol{\rm \Lambda}\boldsymbol{\rm \Theta}\boldsymbol{\rm h}_{{\rm r}})|^2{+}\sigma^2\|\boldsymbol{\rm w}^H\boldsymbol{{\rm H}}\boldsymbol{\rm \Lambda}\boldsymbol{\rm \Theta}\|^2{+}\delta^2)\times\\
&(Q^{-1}(t^*){+}\sqrt{I})\geq \delta^2Q^{-1}(P_f^{\max}){+}\delta^2\sqrt{I}.
\end{split}
\end{equation}
{\color{black}Based on the above transformation of the outer and inner loops,} we can effectually deal with the challenge that the Q function is neither convex nor concave by the inverse of the Q function. However, the above transformation still does not transform the problem (\ref{eq44}) into a convex problem, thus we consider seeking other methods. Furthermore, since the positive and negative properties of the inverse of the Q function are unknown, the constraint cannot be transformed into a convex constraint if the term $p|\boldsymbol{\rm w}^H(\boldsymbol{{\rm h}}_{\rm d}+\boldsymbol{\rm H}\boldsymbol{\rm \Lambda}\boldsymbol{\rm \Theta}\boldsymbol{\rm h}_{{\rm r}})|^2{+}\sigma^2\|\boldsymbol{\rm w}^H\boldsymbol{{\rm H}}\boldsymbol{\rm \Lambda}\boldsymbol{\rm \Theta}\|^2{+}\delta^2$ is transformed into {\color{black} the convex function or the concave function}. {\color{black}Thus, the one-stage beamforming optimization algorithm} can not be applied here. To overcome this problem, we adopt an AO algorithm to transform the constraints into linear constraints, which can decouple each variable so that multiple variables can be optimized separately.

\subsubsection{Receive Beamforming Vector Optimization}
In this section, we optimize the receive beamforming vector for the given phase shift and the amplification factor. Then, problem (\ref{eq44}) can be reduced to 
\begin{equation}\label{eq48}
\begin{split}
\textrm{\textbf{Find}} ~~~&\{\boldsymbol{\rm w}\}\\
s.t.~&{C'_3},\bar C'_4.
\end{split}
\end{equation}
It is worth noting that the equality constraint $C'_3$ is a non-convex constraint. Furthermore, since the positive and negative properties of the inverse of the Q function in $\bar C'_4$ cannot be determined, $\bar C'_4$ remains non-convex even if the term $p|\boldsymbol{\rm w}^H(\boldsymbol{{\rm h}}_{\rm d}+\boldsymbol{\rm H}\boldsymbol{\rm \Lambda}\boldsymbol{\rm \Theta}\boldsymbol{\rm h}_{{\rm r}})|^2$ has been a convex function with respect to $\boldsymbol{\rm w}$.  To this end, we introduce the variable $\boldsymbol{\rm W}=\boldsymbol{\rm w}\boldsymbol{\rm w}^H$.  Then, $\bar C'_4$ can be rewritten as
\begin{equation}\label{eq49}
\begin{split}
&\hat C'_4:(p{\rm Tr}((\boldsymbol{{\rm h}}_{\rm d}+\boldsymbol{\rm H}\boldsymbol{\rm \Lambda}\boldsymbol{\rm \Theta}\boldsymbol{\rm h}_{{\rm r}})^H\boldsymbol{\rm W}(\boldsymbol{{\rm h}}_{\rm d}+\boldsymbol{\rm H}\boldsymbol{\rm \Lambda}\boldsymbol{\rm \Theta}\boldsymbol{\rm h}_{{\rm r}}))\\
&+\sigma^2{\rm Tr}((\boldsymbol{{\rm H}}\boldsymbol{\rm \Lambda}\boldsymbol{\rm \Theta})^H\boldsymbol{\rm W}\boldsymbol{{\rm H}}\boldsymbol{\rm \Lambda}\boldsymbol{\rm \Theta})+\delta^2)(Q^{-1}(t^*)+\sqrt{I})\\
&\geq \delta^2Q^{-1}(P_f^{\max})+\delta^2\sqrt{I}.
\end{split}
\end{equation}
We can see that whether the term $Q^{-1}(t^*)+\sqrt{I}$ is positive or negative, $\hat C'_4$ is a convex constraint. Similarly, the equation constraint $\boldsymbol{\rm W}=\boldsymbol{\rm w}\boldsymbol{\rm w}^H$ can be replaced with (\ref{eq38}). Thus, problem (\ref{eq48}) can be transformed into the following convex problem
\begin{equation}\label{eq50}
\begin{split}
&\textrm{\textbf{Find}} ~~~\{\boldsymbol{\rm w},\boldsymbol{\rm W}\}\\
s.t.~&\hat C'_4,\bar C'_3\!:\!{\rm Tr}(\boldsymbol{\rm W}){=}1,\boldsymbol{\rm W}{\succeq}\boldsymbol{0},\\
&C'_{5a}\!:\!\begin{bmatrix}
\boldsymbol{\rm W}& \boldsymbol{\rm w}\\  
\boldsymbol{\rm w}^H & 1\\ 
\end{bmatrix}\!\! {\succeq} \boldsymbol{0},\\
&\bar C'_{5b}:{\rm Tr}(\boldsymbol{\rm W}){\leq} {-}\|\boldsymbol{\rm w}^{(l)}\|^2{+}2{\rm Tr}((\boldsymbol{\rm w}^{(l)})^H\boldsymbol{\rm w}).
\end{split}
\end{equation}
\subsubsection{Amplification Factor and Phase Shift Optimization}
In this subsection, we optimize the amplification factor and the phase shift for the given receive beamforming vector. Thus, problem (\ref{eq44}) can be reduced to
\begin{equation}\label{eq51}
\begin{split}
\textrm{\textbf{Find}} ~~~&\{\boldsymbol{\rm \Theta},\rho_{n}\}\\
s.t.~&{C'_2},\bar C'_4.
\end{split}
\end{equation}
Since the matrices $\boldsymbol{\rm \Lambda}$ and $\boldsymbol{\rm \Phi}$ are always coupled,  we introduce a new variable $\boldsymbol{\rm v}=[v_1,\cdots,v_n,\cdots,v_N]^T$, where $v_n=\rho_{n}e^{j\theta_n}$. Then, we have the following transformations
$p|\boldsymbol{\rm w}^H(\boldsymbol{{\rm h}}_{\rm d}+\boldsymbol{\rm H}\boldsymbol{\rm \Lambda}\boldsymbol{\rm \Theta}\boldsymbol{\rm h}_{{\rm r}})|^2=p\boldsymbol{\rm\bar v}^H\boldsymbol{\rm H}_{\rm 1}\boldsymbol{\rm\bar v}=p{\rm Tr}(\boldsymbol{\rm\bar V}\boldsymbol{\rm H}_{\rm 1})$,
$\sigma^2\|\boldsymbol{\rm w}^H\boldsymbol{{\rm H}}\boldsymbol{\rm \Lambda}\boldsymbol{\rm \Theta}\|^2=\sigma^2\boldsymbol{\rm\bar v}^H\boldsymbol{\rm H}_2\boldsymbol{\rm H}_2^H\boldsymbol{\rm\bar v}=\sigma^2{\rm Tr}(\boldsymbol{\rm H}_2^H\boldsymbol{\rm\bar V}\boldsymbol{\rm H}_2)$,
$p\|\boldsymbol{\rm \Lambda}\boldsymbol{\rm \Theta}\boldsymbol{\rm h}_{\rm r}\|^2=p\boldsymbol{\rm\bar v}^H\boldsymbol{\rm H}_{\rm r}\boldsymbol{\rm H}_{\rm r}^H\boldsymbol{\rm\bar v}=p{\rm Tr}(\boldsymbol{\rm H}_{\rm r}^H\boldsymbol{\rm\bar V}\boldsymbol{\rm H}_{\rm r})$,
 and $\sigma^2\|\boldsymbol{\rm \Lambda}\boldsymbol{\rm \Theta}\|_F^2={\rm Tr}(\boldsymbol{\rm\bar V}{\rm diag}([\boldsymbol{1}_{N\times 1};0]))$,
where $\boldsymbol{\rm\bar V}=\boldsymbol{\rm\bar v}\boldsymbol{\rm\bar v}^H$ and $\boldsymbol{\rm\bar v}^H\in\mathbb{C}^{1\times (N+1)}=[\boldsymbol{\rm v}^H,1]$, $\boldsymbol{\rm H}_{\rm 1}=\boldsymbol{\rm h}_{\rm 1}\boldsymbol{\rm h}_{\rm 1}^H$, $\boldsymbol{\rm h}_{\rm 1}\in\mathbb{C}^{(N+1)\times 1}=[{\rm diag}(\boldsymbol{\rm w}^H\boldsymbol{\rm H})\boldsymbol{\rm h}_{\rm r}; \boldsymbol{\rm w}^H\boldsymbol{\rm h}_{\rm d}]$, $\boldsymbol{\rm H}_2 \in\mathbb{C}^{(N+1)\times N}=[{\rm diag}(\boldsymbol{\rm w}^H\boldsymbol{\rm H}); \boldsymbol{\rm 0}_{1\times N}
]$. $\boldsymbol{\rm H}_{\rm r}\in\mathbb{C}^{(N+1)\times N}=[{\rm diag}(\boldsymbol{\rm h}_{\rm r}); \boldsymbol{\rm 0}_{1\times N}]$. Then, $\bar C'_4$ can be rewritten as
$\hat C'_4:(p{\rm Tr}(\boldsymbol{\rm\bar V}\boldsymbol{\rm H}_{\rm 1})+\sigma^2{\rm Tr}(\boldsymbol{\rm H}_2^H\boldsymbol{\rm\bar V}\boldsymbol{\rm H}_2)+\delta^2)(Q^{-1}(t^*)+\sqrt{I})\geq \delta^2Q^{-1}(P_f^{\max})+\delta^2\sqrt{I}$.
As a result, problem (\ref{eq51}) can be transformed into the following problem
\begin{equation}\label{eq54}
\begin{split}
&\textrm{\textbf{Find}} ~~~\{\boldsymbol{\rm\bar v},\boldsymbol{\rm\bar V}\}\\
s.t.~&\hat C'_4,{\bar C'_2}:{\rm Pr}(\mathcal{H}_1)t^*(p{\rm Tr}(\boldsymbol{\rm H}_{\rm r}^H\boldsymbol{\rm\bar V}\boldsymbol{\rm H}_{\rm r})\\
&+\sigma^2{\rm Tr}(\boldsymbol{\rm\bar V}{\rm diag}([\boldsymbol{1}_{N\times 1};0])))\leq P_{\rm RIS}^{\max},\\
&C'_5:[\boldsymbol{{\rm\bar V}}]_{N+1,N+1}=1,\boldsymbol{{\rm\bar V}}\succeq \boldsymbol{0},~~C'_6:\boldsymbol{{\rm\bar V}}= \boldsymbol{{\rm\bar v}}\boldsymbol{{\rm\bar v}}^H.
\end{split}
\end{equation}
The equation constraint $\boldsymbol{{\rm\bar V}}= \boldsymbol{{\rm\bar v}}\boldsymbol{{\rm\bar v}}^H$ can be equivalently rewritten as the following constraints
\begin{equation}\label{eq55}
\begin{split}
	&C'_{6a}:\left[                 
	\begin{array}{cc}   
		\boldsymbol{\rm\bar V}& \boldsymbol{\rm\bar v}\\  
		\boldsymbol{\rm\bar v}^H & 1\\  
	\end{array}
	\right] \succeq \boldsymbol{0},\\
	&C'_{6b}:{\rm Tr}(\boldsymbol{\rm\bar V})\leq -\|\boldsymbol{\rm\bar v}^{(l)}\|^2+2{\rm Tr}((\boldsymbol{\rm\bar v}^{(l)})^H\boldsymbol{\rm\bar v}). 
\end{split}
\end{equation}
Thus, problem (\ref{eq54}) can be transformed into the following convex problem
\begin{equation}\label{eq56}
\begin{split}
\textrm{\textbf{Find}} ~~~&\{\boldsymbol{\rm\bar v},\boldsymbol{\rm\bar V}\}\\
s.t.~&{\bar C'_2},\hat C'_4, C'_5,C'_{6a},C'_{6b}.
\end{split}
\end{equation}
\subsubsection{Outer-loop Optimization}
In this subsection, the remaining task is to find the auxiliary variable $t$ in the outer loop. In particular, for the given variables $\boldsymbol{\rm \Theta},\boldsymbol{\rm w},\rho_{n},\epsilon$, we can first determine its the upper bound $t_{\rm max}$ and the lower bound $t_{\min}$, and then apply the bisection method to find the optimal $t$. For a given $t=\frac{t_{\min}+t_{\rm max}}{2}$, when the problems (\ref{eq50}) and (\ref{eq56}) are feasible and thus $\mathcal{S}\not=\emptyset$, then $t$ will be increase by $t_{\min}=t$. When the problems (\ref{eq50}) and (\ref{eq56}) are infeasible and thus $\mathcal{S}=\emptyset$, then $t$ will be decrease by $t_{\max}=t$.
%\begin{spacing}{1.00}
%	\floatname{algorithm}{Algorithm}
%	\renewcommand{\algorithmicrequire}{\textbf{Input:}}
%	\renewcommand{\algorithmicensure}{\textbf{Output:}}
%	\begin{algorithm}[!t]
%		\small
%		\caption{Two-stage Beamforming Optimization Algorithm}
%		\begin{algorithmic}[1]
%			\State Initialize system parameters: $M$, $N$, $P_f^{\max}$,  $\delta^2$, $\sigma^2$, $P_{\rm RIS}^{\max}$, ${\rm Pr}(\mathcal{H}_1)$, $p$, $I$;
%			\State Set the maximum iteration number $L^{\max}$ and the convergence accuracy $\epsilon_1$, set the initial iteration index $ll=0$;
%			\State Set the upper bound $t_{\max}$ and the lower bound $t_{\min}$ of the auxiliary variable $t$;
%			\While{$t_{\max}-t_{\min}>\epsilon_2$}
%			\State Set $t=\frac{t_{\max}+t_{\min}}{2}$;
%			\Repeat
%			\State Solve the convex feasibility problem (\ref{eq50}), obtain $\boldsymbol{\rm w}$ and  $\boldsymbol{\rm W}$;
%			\Until{Convergence}
%			\Repeat
%			\State Solve the convex feasibility problem (\ref{eq56}), obtain $\boldsymbol{\rm\bar v}$ and  $\boldsymbol{\rm\bar V}$;
%			\Until{Convergence}
%			\State find $\mathcal{S}$;
%			\If{Problems (\ref{eq50}) and (\ref{eq56}) feasible and $\mathcal{S}\not=\emptyset$}
%			\State $t_{\min}=t$;
%			\Else
%			\State $t_{\max}=t$;
%			\EndIf
%			\EndWhile
%			\State Output $t^*=\frac{t_{\max}+t_{\min}}{2}$	.	
%		\end{algorithmic}
%	\end{algorithm}
%\end{spacing}
\subsection{The Upper and Lower Bounds of The Detection Probability}
First, we observe the optimization problem (\ref{eq24}). It is obvious that $C'_2$ is {\color{black}active. That is,} the equality always holds. If the equality does not hold, the detection probability can continue to increase until the equality holds. Then, we can obtain
$P_d^{\rm act}=\frac{P_{\rm RIS}^{\max}}{{\rm Pr}(\mathcal{H}_1)(p\|\boldsymbol{\rm \Lambda}\boldsymbol{\rm \Theta}\boldsymbol{\rm h}_{\rm r}\|^2+\sigma^2\|\boldsymbol{\rm \Lambda}\boldsymbol{\rm \Theta}\|_F^2)}$.
Then, we have the following proposition.

\textit{\textbf{Proposition 5:}} Given the amplification factor, the upper and lower bounds of the detection probability are given by
\begin{equation}\label{eq58}
\begin{split}
\frac{P_{\rm RIS}^{\max}}{N\lambda_{\max}(\rho_{n})}\leq P_d^{\rm act}\leq \frac{P_{\rm RIS}^{\max}}{N\lambda_{\min}(\rho_{n})},
\end{split}
\end{equation}
where $\lambda_{\min}$ and $\lambda_{\max}$ are the smallest and largest elements of the principal diagonal of the matrix ${\rm Pr}(\mathcal{H}_1)(p{\rm diag}(\boldsymbol{\rm h}_{\rm r}^H)\boldsymbol{\rm \Lambda}^H\boldsymbol{\rm \Lambda}{\rm diag}(\boldsymbol{\rm h}_{\rm r})+\sigma^2\boldsymbol{\rm \Lambda}^H\boldsymbol{\rm \Lambda})$, respectively.
\begin{proof}
	Please refer to Appendix D.
\end{proof}

\section{\color{black}The Number Configuration for the Passive/Active RIS-aided Spectrum Sensing}
In this section, we investigate the number configuration (see, e.g., \cite{i4} and \cite{i4a}) by analyzing how many reflecting elements are needed to make the detection probability close to 1. To facilitate our analysis, the ST is equipped with a single antenna, i.e., $\boldsymbol{\rm H}\in \mathbb{C}^{M\times N}\rightarrow\boldsymbol{\rm h}\in \mathbb{C}^{1\times N}$. Moreover, we ignore the direct link, and the amplification factor of each reflecting element is the same, i.e., $\rho_{n}=\rho,\forall n$ \cite{i5}.
\subsection{\color{black}The Number Configuration for Passive RIS-aided Spectrum Sensing}
Then, the detection probability maximization problem for the passive RIS can be reduced to
\begin{equation}\label{eq62}
\begin{array}{l}
\max\limits_{\mbox{\scriptsize$\begin{array}{c} 
		\boldsymbol{\rm \Theta},\epsilon
		\end{array}$}} 
Q\left(\left(\frac{\epsilon}{\delta^2}-\dfrac{p|\boldsymbol{\rm h}\boldsymbol{\rm \Theta}\boldsymbol{\rm h}_{{\rm r}}|^2}{\delta^2}-1\right)\sqrt{\frac{I}{(1+\dfrac{p|\boldsymbol{\rm h}\boldsymbol{\rm \Theta}\boldsymbol{\rm h}_{{\rm r}}|^2}{\delta^2})^2}}\right)\\
s.t.~{C_1}: Q\left(\left(\frac{\epsilon}{\delta^2}-1\right)\sqrt{I}\right)\leq P_f^{\max},~{C_2}:0\leq\theta_{n}<2\pi.
\end{array}
\end{equation}
\subsubsection{Detection Threshold Optimization }
Based on the monotonicity of the Q function and  $C_1$, we can design the detection threshold as $\epsilon^* = \frac{\delta^2 Q^{-1}(P_f^{\max})}{\sqrt{I}}+\delta^2$ to meet the maximum probability of false alarm.
\subsubsection{Phase-shift Optimization }
The Q function is a monotonically non-increasing function,  (\ref{eq62}) can be transformed into
\begin{equation}\label{eq64}
\begin{split}
\max\limits_{\mbox{\scriptsize$\begin{array}{c} 
		\boldsymbol{\rm \Theta}
		\end{array}$}} 
&~\dfrac{p|\boldsymbol{\rm h}\boldsymbol{\rm \Theta}\boldsymbol{\rm h}_{{\rm r}}|^2}{\delta^2}\\
s.t.~&{C_2}.
\end{split}
\end{equation}
According to the above optimization problem, it is obvious that the optimal phase shift can be obtained as
$\theta_{n}^*=\arg([\boldsymbol{\rm h}]_n)-\arg([\boldsymbol{\rm h}_{\rm r}]_n)$.
Then, we substitute the optimal phase shift into the SNR expression, which can be rewritten as
\begin{equation}\label{eq66}
\begin{split}
&\dfrac{p|\boldsymbol{\rm h}\boldsymbol{\rm \Theta}\boldsymbol{\rm h}_{{\rm r}}|^2}{\delta^2}\overset{(a)}{=}\frac{p\left|\sum\limits_{n=1}^N|h_n||h_{{\rm r},n}|\right|^2}{\delta^2}\overset{(b)}{\geq} \frac{pN^2|h|^2|h_{{\rm r}}|^2}{\delta^2},
\end{split}
\end{equation}
where $(a)$ utilizes the optimal design of $\boldsymbol{\rm \Theta}$, $(b)$ utilizes $|h|=\min\{|h_n|\}$ and $|h_{{\rm r}}|=\min\{|h_{{\rm r},n}|\}$. $h_n$ and $h_{{\rm r},n}$ are the $n$-th elements of $\boldsymbol{\rm h}$ and $\boldsymbol{\rm h}_{{\rm r}}$.
Then, we substitute the detection threshold and the SNR expression into the detection probability, which can be rewritten as
\begin{equation}\label{eq67}
\begin{split}
Q\left(\frac{\delta^2 Q^{-1}(P_f^{\max})+\delta^2\sqrt{I}}{pN^2|h|^2|h_{{\rm r}}|^2+\delta^2}-\sqrt{I}\right).
\end{split}
\end{equation}
Based on the Three-Sigma Rule for Gaussian distributions, we have
${\rm Pr}(\mu-3\sigma \leq X\leq \mu+3\sigma)\approx 0.9974$ and ${\rm Pr}(X\geq \mu-3\sigma)\approx 0.9987$, where $X\sim \mathcal{N}(\mu, \sigma^2)$. Based on Appendix C, we have $Q(-3)\approx 0.9987$. Thus, when the following inequality holds, the detection probability is approximately 1
 \begin{equation}\label{eq68}
 \begin{split}
\frac{\delta^2 Q^{-1}(P_f^{\max})+\delta^2\sqrt{I}}{pN^2|h|^2|h_{{\rm r}}|^2+\delta^2}-\sqrt{I}\leq -3.
 \end{split}
 \end{equation}
Since the maximum tolerable probability of false alarm $P_f^{\max}$ is usually a small constant not exceeding 0.1, thus $Q^{-1}(P_f^{\max})>0$ holds. Meanwhile, $\sqrt{I}\gg 3$. Then, we have
\begin{equation}\label{eq69}
\begin{split}
N\geq \sqrt{\frac{\delta^2 Q^{-1}(P_f^{\max})+\delta^2\sqrt{I}}{(\sqrt{I}-3)p|h|^2|h_{{\rm r}}|^2}-\frac{\delta^2}{p|h|^2|h_{{\rm r}}|^2}}.
\end{split}
\end{equation}
\subsection{\color{black}The Number Configuration for Active RIS-aided Spectrum Sensing}
The detection probability maximization problem for the active RIS can be reduced to
\begin{equation}\label{eq70}
\begin{split}
\max\limits_{\mbox{\scriptsize$\begin{array}{c} 
		\boldsymbol{\rm \Theta},\rho,\epsilon
		\end{array}$}} 
&Q\left(\left(\frac{\epsilon}{\sigma^2\|\boldsymbol{{\rm h}}\boldsymbol{\rm \Lambda}\boldsymbol{\rm \Theta}\|^2+\delta^2}-\gamma^{\rm act}-1\right)\frac{\sqrt{I}}{1+\gamma^{\rm act}}\right)\\
s.t.~{C'_1}: &P_f^{\rm act}\leq P_f^{\max},\\
{C'_2}:&{\rm Pr}(\mathcal{H}_1)P_d^{\rm act}(p\|\boldsymbol{\rm \Lambda}\boldsymbol{\rm \Theta}\boldsymbol{\rm h}_{\rm r}\|^2+\sigma^2\|\boldsymbol{\rm \Lambda}\boldsymbol{\rm \Theta}\|_F^2)\leq P_{\rm RIS}^{\max},
\end{split}
\end{equation}
where
$\gamma^{\rm act}=p|\boldsymbol{\rm h}\boldsymbol{\rm \Lambda}\boldsymbol{\rm \Theta}\boldsymbol{\rm h}_{{\rm r}}|^2/(\sigma^2\|\boldsymbol{{\rm h}}\boldsymbol{\rm \Lambda}\boldsymbol{\rm \Theta}\|^2+\delta^2)$.
First, we can design the detection threshold as (\ref{eq20}) to meet the maximum probability of false alarm. Since the detection probability in $C'_2$ is very tricky for us to deal with, we apply a one-stage algorithm to reduce $C'_2$ to $\bar C'_2:{\rm Pr}(\mathcal{H}_1)(p\|\boldsymbol{\rm \Lambda}\boldsymbol{\rm \Theta}\boldsymbol{\rm h}_{\rm r}\|^2+\sigma^2\|\boldsymbol{\rm \Lambda}\boldsymbol{\rm \Theta}\|_F^2)\leq P_{\rm RIS}^{\max}$. Then, (\ref{eq70}) can be transformed into
\begin{equation}\label{eq72}
\begin{split}
\max\limits_{\mbox{\scriptsize$\begin{array}{c} 
		\boldsymbol{\rm \Theta},\rho
		\end{array}$}} 
&p|\boldsymbol{\rm h}\boldsymbol{\rm \Lambda}\boldsymbol{\rm \Theta}\boldsymbol{\rm h}_{{\rm r}}|^2+\sigma^2\|\boldsymbol{{\rm h}}\boldsymbol{\rm \Lambda}\boldsymbol{\rm \Theta}\|^2+\delta^2\\
s.t.~&{\bar C'_2}.
\end{split}
\end{equation}
According to the above optimization problem, it is obvious that the optimal phase shift can be obtained as $\theta_{n}^*=\arg([\boldsymbol{\rm h}]_n)-\arg([\boldsymbol{\rm h}_{\rm r}]_n)$.
Then, we substitute the optimal phase shift into the objective function and $C'_2$, which can be rewritten as
\begin{equation}\label{eq74}
\begin{array}{l}
p|\boldsymbol{\rm h}\boldsymbol{\rm \Lambda}\boldsymbol{\rm \Theta}\boldsymbol{\rm h}_{{\rm r}}|^2+\sigma^2\|\boldsymbol{{\rm h}}\boldsymbol{\rm \Lambda}\boldsymbol{\rm \Theta}\|^2+\delta^2\\
=p\left|\boldsymbol{\rm h}\boldsymbol{\rm \Lambda}\boldsymbol{\rm \Theta}\boldsymbol{\rm h}_{{\rm r}}\right|^2+\sigma^2|h|^2\sum\limits_{n=1}^N\rho^2+\delta^2\\
%\end{array}
%\end{equation}
%\begin{equation}\notag
%\begin{array}
\overset{(a)}{=} p|\sum\limits_{n=1}^N\rho|h||h_{\rm r}||^2+\sigma^2|h|^2N\rho^2+\delta^2\\
\overset{(b)}{\geq}p|h|^2|h_{\rm r}|^2N^2\rho^2+\sigma^2|h|^2N\rho^2+\delta^2,
\end{array}
\end{equation}
\begin{equation}\label{eq75}
\begin{split}
&p\|\boldsymbol{\rm \Lambda}\boldsymbol{\rm \Theta}\boldsymbol{\rm h}_{\rm r}\|^2+\sigma^2\|\boldsymbol{\rm \Lambda}\boldsymbol{\rm \Theta}\|_F^2=p|h_{\rm r}|^2N\rho^2+\sigma^2N\rho^2,
\end{split}
\end{equation}
where $(a)$ utilizes the optimal design of $\boldsymbol{\rm \Theta}$, $(b)$ utilizes $|h|=\min\{|h_n|\}$ and $|h_{{\rm r}}|=\min\{|h_{{\rm r},n}|\}$. $h_n$ and $h_{{\rm r},n}$ are the $n$-th elements of $\boldsymbol{\rm h}$ and $\boldsymbol{\rm h}_{{\rm r}}$.
Then, (\ref{eq72}) can be transformed into
\begin{equation}\label{eq76}
\begin{split}
\max\limits_{\mbox{\scriptsize$\begin{array}{c} 
		\rho
		\end{array}$}} 
&~p|h|^2|h_{\rm r}|^2N^2\rho^2+\sigma^2|h|^2N\rho^2+\delta^2\\
s.t.~&{\hat C'_2}:{\rm Pr}(\mathcal{H}_1)(p|h_{\rm r}|^2N\rho^2+\sigma^2N\rho^2)\leq P_{\rm RIS}^{\max}.
\end{split}
\end{equation}
{\color{black}Since the objective function increases with the increasing $\rho$, it can be maximized} when $\rho$ takes its maximum value. Thus, $\hat C'_2$ is active, the optimal $\rho^*$ is given by
\begin{equation}\label{eq77}
\begin{split}
\rho^*=\sqrt{\frac{P_{\rm RIS}^{\max}}{{\rm Pr}(\mathcal{H}_1)N(p|h_{\rm r}|^2+\sigma^2)}}.
\end{split}
\end{equation}
Then, substituting the phase shift and amplification factor into the detection probability, we have
\begin{equation}\label{eq78}
\begin{split}
P_d^{\rm act}=Q\left(\frac{\delta^2 Q^{-1}(P_f^{\max})+\delta^2\sqrt{I}}{\frac{pN|h|^2|h_{\rm r}|^2P_{\rm RIS}^{\max}+\sigma^2|h|^2P_{\rm RIS}^{\max}}{{\rm Pr}(\mathcal{H}_1)(p|h_{\rm r}|^2+\sigma^2)}+\delta^2}-\sqrt{I}\right).
\end{split}
\end{equation}
Based on  the Three-Sigma Rule for Gaussian distributions, we have
\begin{equation}\label{eq79}
\begin{array}{l}
\frac{\delta^2 Q^{-1}(P_f^{\max})+\delta^2\sqrt{I}}{\frac{pN|h|^2|h_{\rm r}|^2P_{\rm RIS}^{\max}+\sigma^2|h|^2P_{\rm RIS}^{\max}}{{\rm Pr}(\mathcal{H}_1)(p|h_{\rm r}|^2+\sigma^2)}+\delta^2}-\sqrt{I}\leq -3.\\
\end{array}
\end{equation}
Thus, when the number of reflecting elements meets the following condition, the detection probability is approximately 1
\begin{equation}\label{eq80}
\begin{split}
N&\geq \left[\frac{{\rm Pr}(\mathcal{H}_1)(p|h_{\rm r}|^2{+}\sigma^2)(\delta^2 Q^{-1}(P_f^{\max}){+}\delta^2\sqrt{I})}{(\sqrt{I}-3)p|h|^2|h_{\rm r}|^2P_{\rm RIS}^{\max}} \right.\\
&\left.-\frac{{\rm Pr}(\mathcal{H}_1)\delta^2(p|h_{\rm r}|^2{+}\sigma^2){+}\sigma^2|h|^2P_{\rm RIS}^{\max}}{p|h|^2|h_{\rm r}|^2P_{\rm RIS}^{\max}} \right]^+.
\end{split}
\end{equation}
\subsection{\color{black}Performance Comparison for passive/active RIS-aided Spectrum Sensing}
We define the number of passive reflecting elements as $N_{\rm pas}$ and the number of active reflecting elements as $N_{\rm act}$, the detection probability for the passive RIS and the active RIS are written as
\begin{equation}\label{eq81}
\begin{split}
&P_d^{\rm act}{=}Q\!\!\left(\!(\frac{\epsilon}{pN_{\rm pas}^2|h|^2|h_{{\rm r}}|^2+\delta^2}{-}1\!)\sqrt{I}\right),\\
&P_d^{\rm act}{=}Q\!\!\left(\!(\frac{\epsilon}{p|h|^2|h_{\rm r}|^2N_{\rm act}^2\rho^2+\sigma^2|h|^2N_{\rm act}\rho^2+\delta^2}{-}1)\sqrt{I}\!\right).
\end{split}
\end{equation}
The direct comparison of detection probability is a {\color{black}challenging task. However, according to} the monotonicity of the Q function, we only need to compare their sum of useful information power and noise power to compare their detection performance, i.e., $pN_{\rm pas}^2|h|^2|h_{{\rm r}}|^2+\delta^2$ and $p|h|^2|h_{\rm r}|^2N_{\rm act}^2\rho^2+\sigma^2|h|^2N_{\rm act}\rho^2+\delta^2$. Thus,  we have the following Proposition.

\textbf{\textit{Proposition 6:}}
The detection probability for active RIS outperforms that for passive RIS if
\begin{equation}\label{eq82}
\begin{split}
\frac{p|h|^2|h_{\rm r}|^2N_{\rm act}P_{\rm RIS}^{\max}+\sigma^2|h|^2P_{\rm RIS}^{\max}}{{\rm Pr}(\mathcal{H}_1)(p|h_{\rm r}|^2+\sigma^2)}>p|h|^2|h_{\rm r}|^2N_{\rm pas}^2.
\end{split}
\end{equation}
Otherwise, the detection probability for passive RIS outperforms that for active RIS.

\textbf{\textit{Corollary:}} The detection probability for active RIS outperforms that for passive RIS if
	\begin{equation}\label{eq85}
	\begin{split}
N_{\rm act}^2+\frac{\sigma^2}{p|h_{\rm r}|^2}N_{\rm act}>N_{\rm pas}^2.
	\end{split}
	\end{equation}
It is worth noting that when (\ref{eq85}) holds, (\ref{eq82}) must hold.

\textbf{\textit{Remark 5:}} It can be seen from Proposition 6 and Corollary that the detection probability of the active RIS is not necessarily better than that of the passive RIS. With the same power budget and the PT is active,  we can easily get $\frac{N_{\rm act}}{N_{\rm pas}}<1$. Thus, when the number of passive reflecting elements is much greater than the number of active reflecting elements or the maximum power of active RIS is quite small, the detection performance of the passive RIS may be better than the active RIS. Meanwhile, when the number of reflecting elements for the active RIS and the passive RIS is the same, the detection performance of the active RIS must be {\color{black}better than that of the passive RIS.}

\section{Simulation Results}
In this section, the effectiveness of the proposed algorithms is evaluated by simulation results. We define the one-stage beamforming optimization algorithm as ``Active RIS 1'' and define the two-stage optimization algorithm as ``Active RIS 2''. Besides,  the proposed passive RIS-aided spectrum sensing is defined as ``Passive RIS'', which is referred to Section III.  Furthermore, ``Without RIS'' is  a traditional spectrum sensing algorithm, there is no RIS to assist sensing and only the direct link. The false alarm and detection probabilities are readily available:
%\begin{equation}\label{eq86}
%\begin{split}
$P_f=Q((\frac{\epsilon}{\delta^2}-1)\sqrt{I})$,
$P_d=Q((\frac{\epsilon}{\delta^2}-\gamma^{\rm no}-1)\sqrt{\frac{I}{(1+\gamma^{\rm no})^2}})$,
%\end{split}
%\end{equation}
where $\gamma^{\rm no}=\frac{p|\boldsymbol{{\rm w}}^H\boldsymbol{{\rm h}}_{\rm d}|^2}{\delta^2}$. 

A two-dimensional coordinate setup measured in meter (m) is considered, where the PT and the RIS are located at $(0, 0)$ m, $(50, 20)$ m, while the ST is deployed at $(500, 0)$ m. {\color{black}The RIS-based channel model includes the large-scale path-loss model and the small-scale multipath fading model. For the large-scale path-loss model, we establish it as $A=A_0(\frac{d}{d_0})^{-\beta}$\cite{i5},} where $A_0=-30$ dB is the path-loss factor at $d_0=1$ m, $d$ is the distance between the transmitter and the receiver.  The path-loss exponents for the PT-RIS link, the RIS-ST link, and PT-ST link are $2$, $2.6$, and $3.5$, respectively. The small-scale fading follows the Rayleigh distribution\footnote{\color{black}Rayleigh fading is a suitable case for scenarios where the RIS is randomly deployed, the location of the RIS cannot be optimized, and therefore the existence of a strong line-of-sight component cannot be guaranteed\cite{i5}.}. Other parameters are: $p\in[5,30]$ dBm, $P_f^{\max}=0.1$, $P_{\rm RIS}^{\max}\in[-30,10]$ dBm, $\delta^2=-70$ dBm, $\sigma^2=-70$ dBm, $M=8$, $N=6$, $t_s = 0.001$ S, $f_s=6$ MHz, and ${\rm Pr}(\mathcal{H}_1)=0.5$.

\subsection{Performance Analysis of Proposed Algorithms}
\begin{figure}%\vspace{-20pt}
	\centering
	\begin{subfigure}{0.49\linewidth}
		\centering
		\includegraphics[width=1\linewidth]{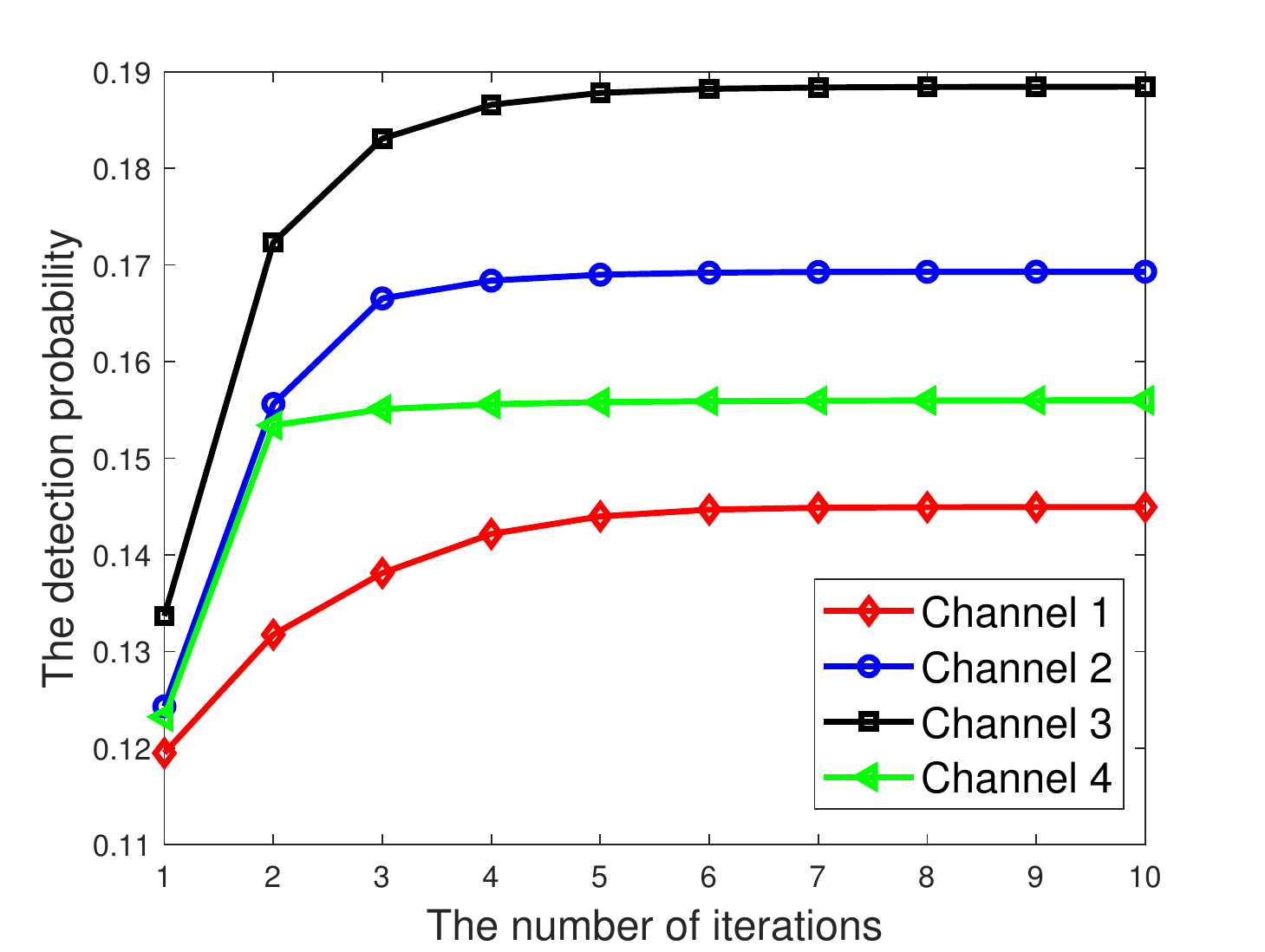}
		\caption{}
		\label{fig3}%文中引用该图片代号
	\end{subfigure}
	\centering
	\begin{subfigure}{0.49\linewidth}
		\centering
		\includegraphics[width=1\linewidth]{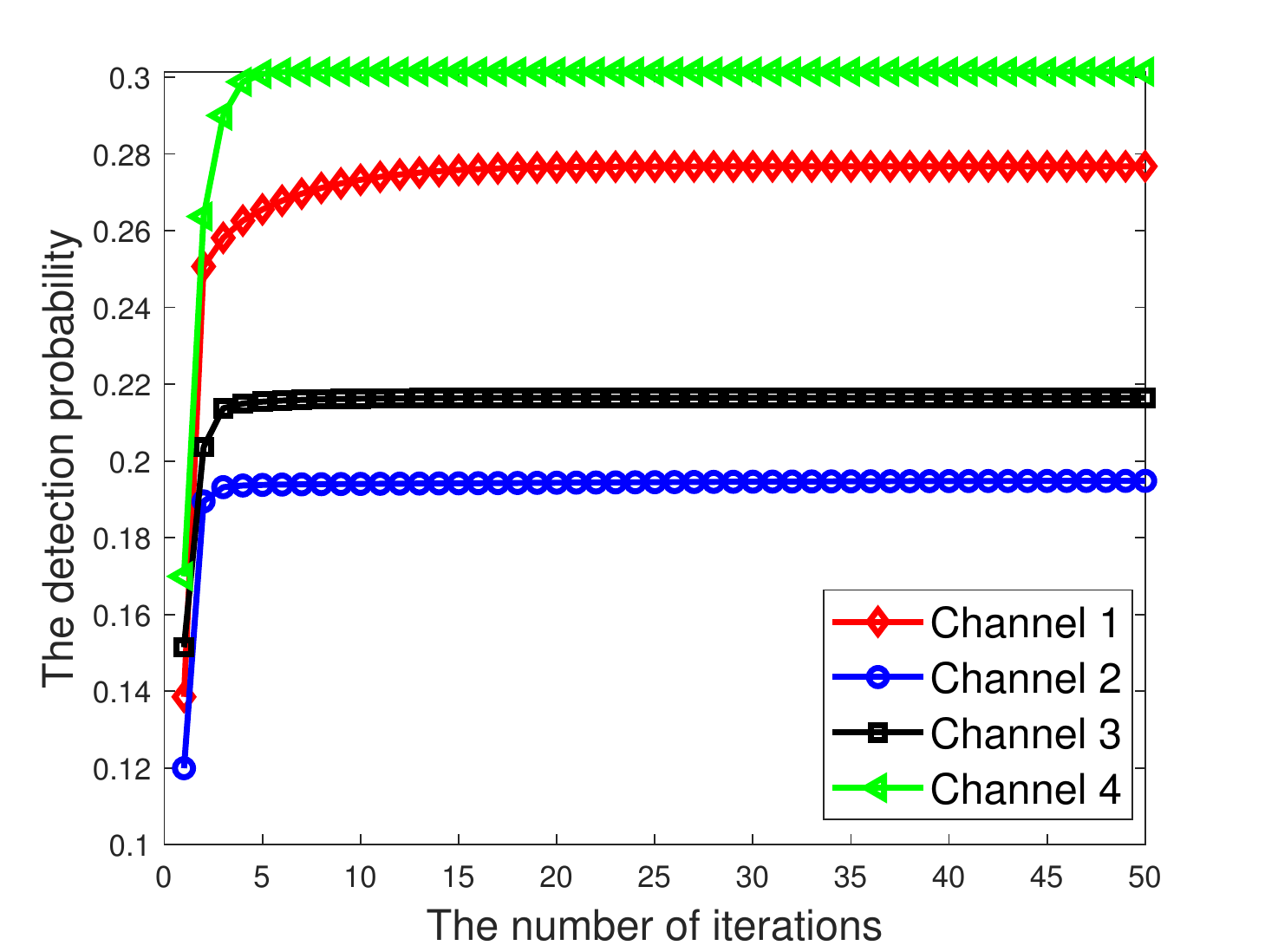}
		\caption{}
		\label{fig4}%文中引用该图片代号
	\end{subfigure}
	\caption{Convergence for proposed algorithms. (a) The active RIS 1. (b)  The active RIS 2.}
	\label{fig34}\vspace{-15pt}
\end{figure}
\begin{figure}\vspace{-15pt}
	\centering
	\begin{subfigure}{0.49\linewidth}
		\centering
		\includegraphics[width=1\linewidth]{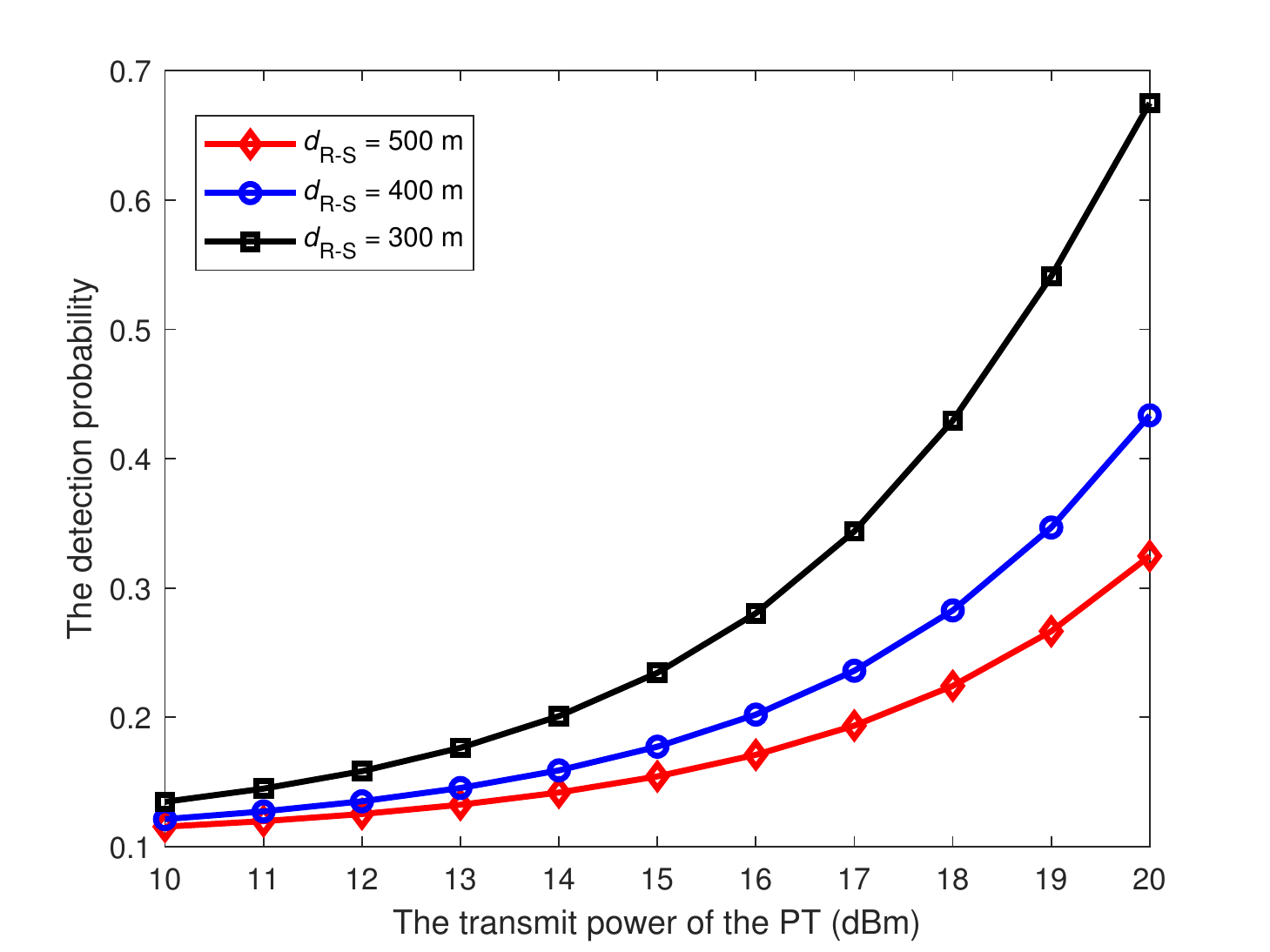}
		\caption{Rayleigh case.}
		\label{fig4-1}%文中引用该图片代号
	\end{subfigure}
	\centering
	\begin{subfigure}{0.49\linewidth}
		\centering
		\includegraphics[width=1\linewidth]{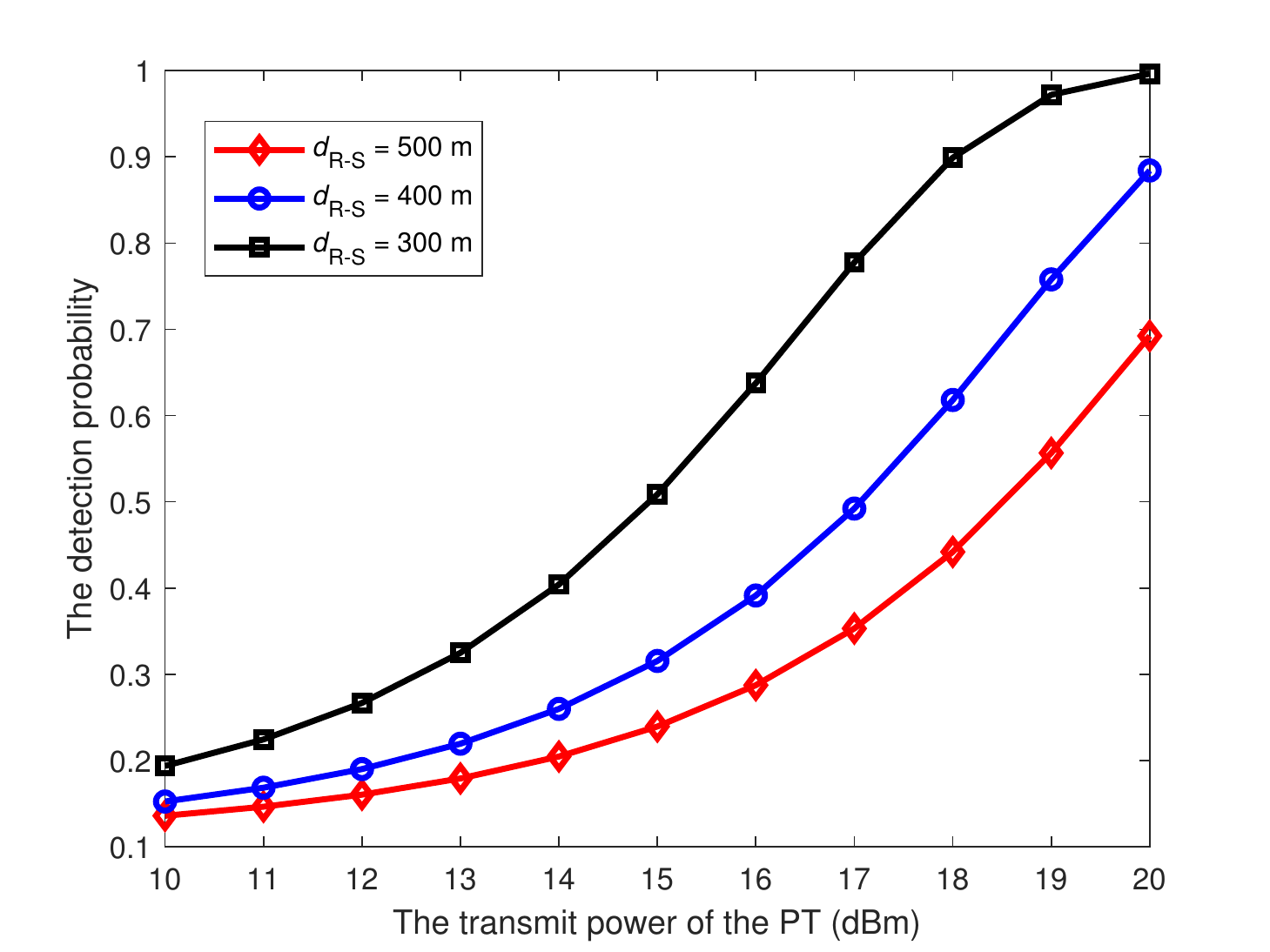}
		\caption{Rician case, Rician factor is 3 dB.}
		\label{fig4-2}
	\end{subfigure}
	\caption{\color{black}The detection probability versus channel conditions and the distance between the RIS and the ST for the passive RIS.}
	\label{fig4-12}
\end{figure}

Fig. \ref{fig34} evaluates the convergence performance of the proposed algorithms for the active RIS 1 and the active RIS 2 under four arbitrarily selected channel realizations, respectively. Specifically,  it is observed from Fig. \ref{fig34}(a) that the active RIS 1 converges to the stable value after ten iterations, which reflects that the active RIS 1 has good convergence performance. Besides, it can be observed from  Fig. \ref{fig34}(b) that after twenty iterations, the active RIS 2 converges to a stable value. Obviously, the number of iterations of the active RIS 2 is higher than that of the active RIS 1. This is because the active RIS 2 is a two-stage optimization algorithm, and the computational complexity of the bisection method for the outer-loop optimization is higher. 
{\color{black}Fig. \ref{fig4-12} gives the detection probability for the passive RIS versus channel conditions and the distance between the RIS and the ST. From Fig. \ref{fig4-12}, we can observe that as the distance between the RIS and the ST increases, the detection probability decreases. Additionally, compared to the Rayleigh case, a higher detection probability can be obtained under the Rician case, which also reflects that the optimization framework proposed in this work can be easily extended to channels with Rician fading.}

\begin{figure}[t]
	%\vspace{-20pt}
	\centering
	\begin{minipage}[t]{0.24\textwidth}
		\centering
		\includegraphics[width=1.7in]{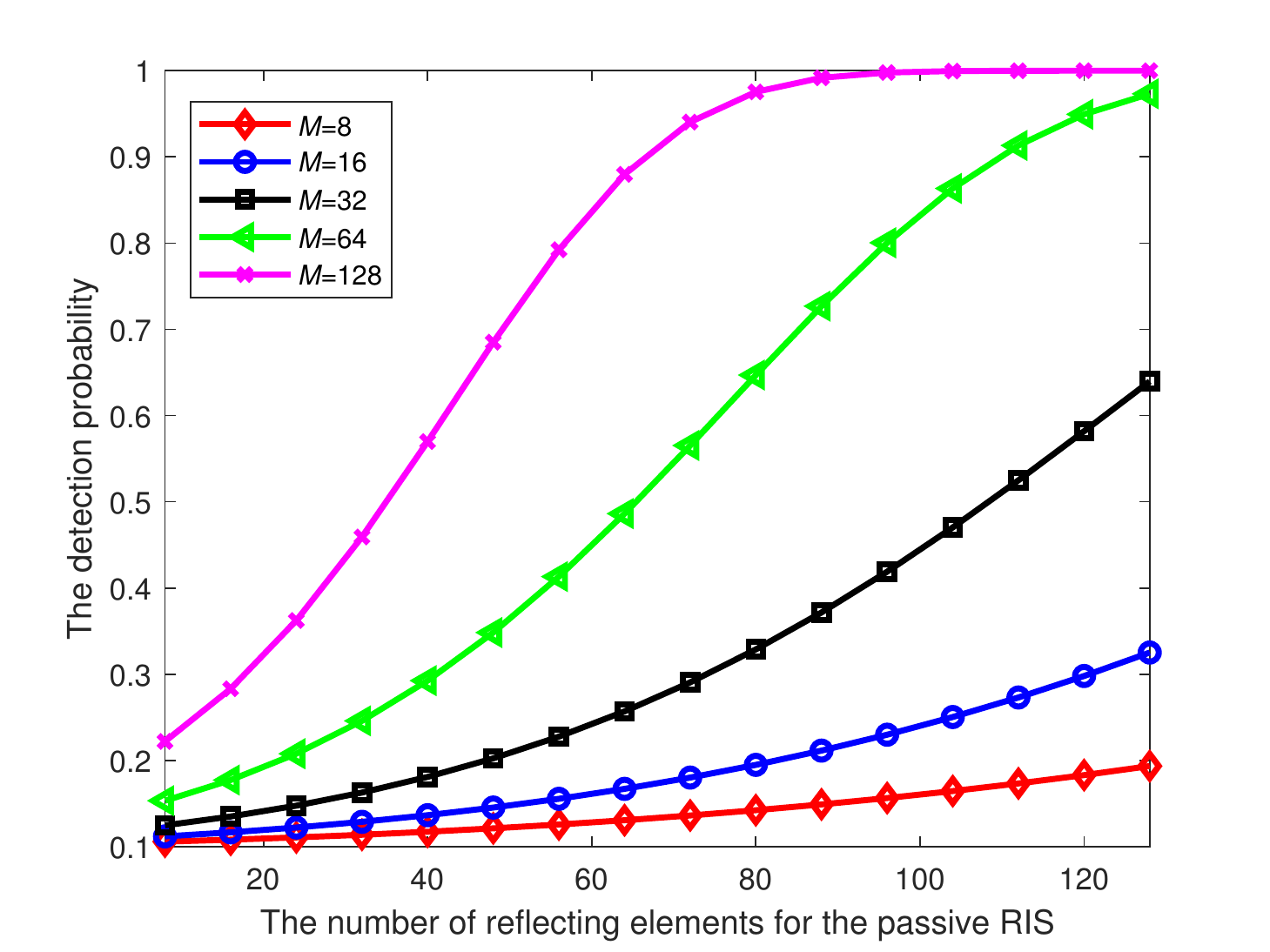}
		\caption{The detection probability versus the number of reflecting elements for the passive RIS.}
		\label{fig5}
	\end{minipage}
	\begin{minipage}[t]{0.24\textwidth}
		\centering
		\includegraphics[width=1.7in]{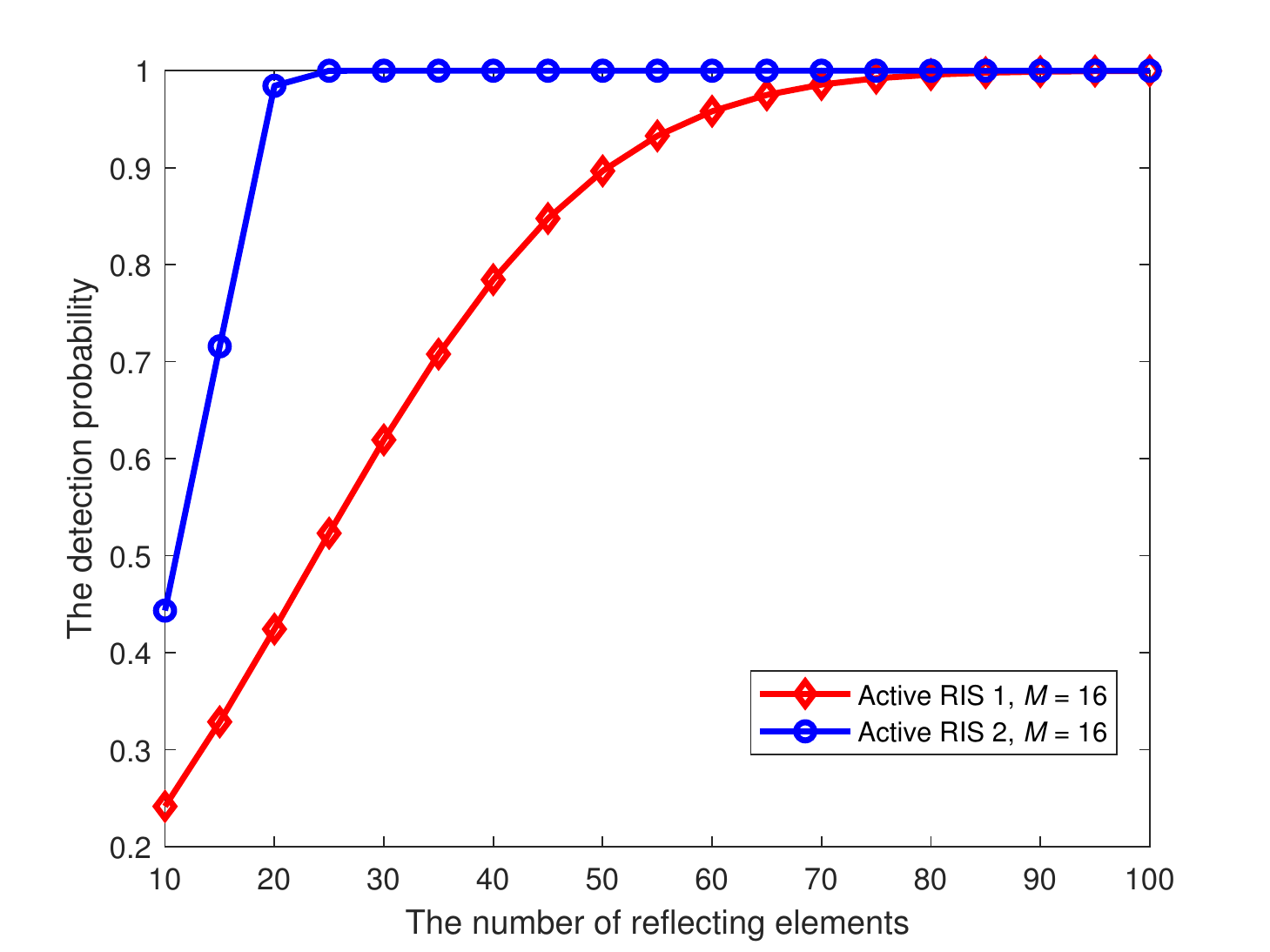}
		\caption{The detection probability versus the number of reflecting elements for the active RIS.}
		\label{fig5-0}
	\end{minipage}
\end{figure}

{\color{black}Fig. \ref{fig5} shows the detection probability versus the number of reflecting elements for the passive RIS. It can be observed that the detection probability increases with the increasing number of reflecting elements. This trend is attributed to the fact that more reflecting elements for the passive RIS can achieve higher passive beamforming gains, thereby improving the SNR performance. According to  Eq. (\ref{eq9}), we can obtain a higher detection probability. Besides, the detection probability increases with the increasing number of antennas for the ST. This is because more antennas at the ST can effectively enhance the received signal strength, resulting in a higher detection performance. Under the corresponding parameter configuration, when $M=128$ holds, approximately $96$ passive reflecting elements are needed to make the detection probability close to 1.}
%shows the detection probability versus the number of reflecting elements for the passive RIS. It can be observed that the detection probability increases with the increasing number of reflecting elements. According to (\ref{eq9}), the detection probability of passive RIS-enhanced spectrum sensing is dependent on the SNR. With more reflecting elements, the passive RIS can achieve finer control over the phase and amplitude of the reflected signals, which enables precise beamforming and signal steering, and allows the passive RIS to concentrate the signal energy towards the ST. Thus, the received signal strength at the ST can be significantly increased, leading to an improved detection probability. On the other hand, the detection probability increases with the increasing number of antennas for the ST. This is because more antennas allow the ST to capture a larger portion of the available signal energy. By spatially combining the received signals, the ST can effectively increase the signal strength, resulting in an improved SNR and a higher detection accuracy. Under the corresponding parameter configuration, when $M=128$, approximately $96$ passive reflecting elements are needed to make the detection probability close to 1. 
Fig. \ref{fig5-0} shows the detection probability versus the number of reflecting elements for the active RIS, where $P_{\rm RIS}^{\max}=-10$ dBm. It can be seen that, compared with the passive RIS, the active RIS only needs a small number of reflecting elements and antennas to make the detection probability close to 1. That is, under the corresponding parameter configuration, when $M=16$, active RIS 1 and active RIS 2 need about $80$ and $25$ reflecting elements to make the detection probability close to 1, respectively.

%\begin{figure}\vspace{-20pt}
%	\centering
%	\includegraphics[width=2in]{fig5.eps}
%	\caption{The detection probability versus the number of reflecting elements for the passive RIS.}
%	\label{fig5}%\vspace{-20pt}
%\end{figure}
%\begin{figure}\vspace{-15pt}
%	\centering
%	\includegraphics[width=2in]{fig5-0.eps}
%	\caption{The detection probability versus the number of reflecting elements for the active RIS.}
%	\label{fig5-0}\vspace{-15pt}
%\end{figure}

\begin{figure}[t]
\vspace{-15pt}
	\centering
	\begin{minipage}[t]{0.24\textwidth}
		\centering
		\includegraphics[width=1.7in]{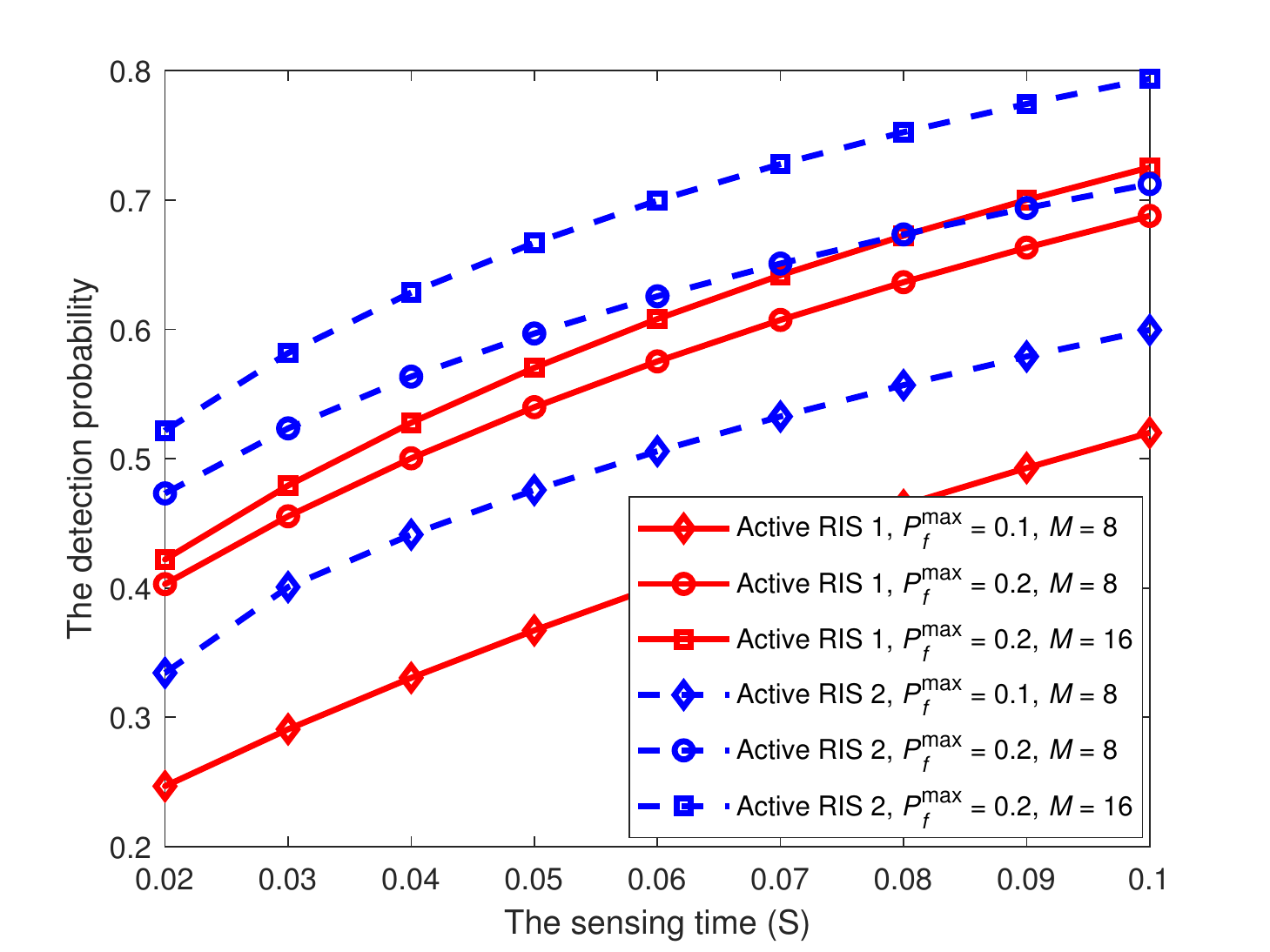}
		\caption{The detection probability versus the sensing time.}
		\label{fig5-1}%\vspace{-20pt}
	\end{minipage}
	\begin{minipage}[t]{0.24\textwidth}
		\centering
		\includegraphics[width=1.7in]{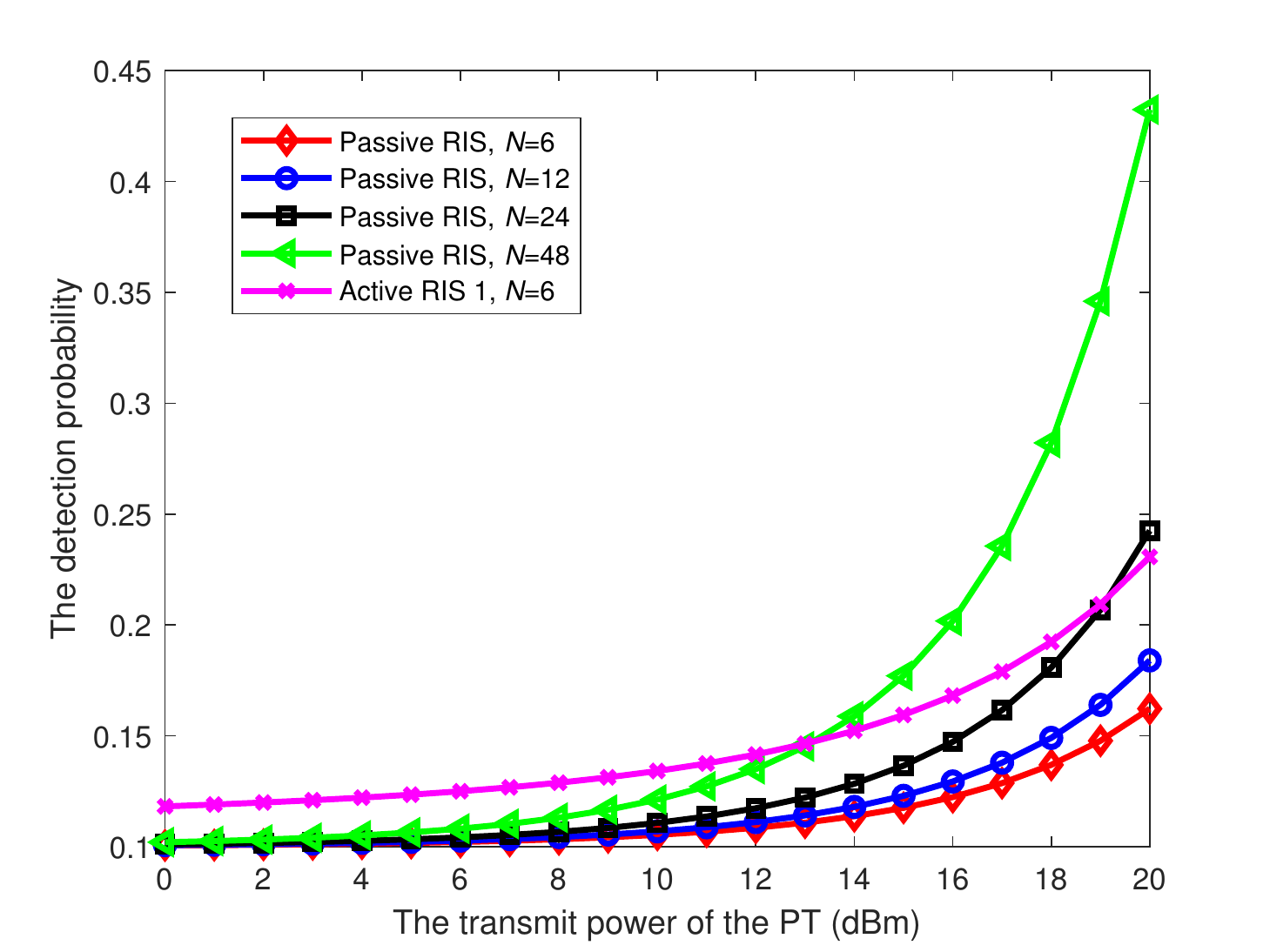}
		\caption{The detection probability versus the transmit power of the PT.}
		\label{fig6}%\vspace{-20pt}
	\end{minipage}
\end{figure}

Fig. \ref{fig5-1} gives the detection probability versus the sensing time for the active RIS under the proposed different algorithms. As the sensing time increases, the detection probability also improves. This is because, with the increase in sensing time, more observation samples can be collected, which provides more information to determine spectrum utilization. Based on the analysis and processing of a larger number of samples, the accuracy and reliability of detection can be enhanced. In addition, the detection probability increases with the increase of the false alarm threshold. This is due to the fact as the tolerated probability of false alarm increases, the detection threshold {\color{black}can be designed to be lower}. This means that the detection becomes more sensitive to signals and more observation samples are classified as target signals, which enhances the ability to detect signals and increases the detection probability. Furthermore, under the same parameter configuration, the detection probability in Active RIS 2 outperforms Active RIS 1. Because Active RIS 1 scales the constraints, which may reduce the feasible region, while Active RIS 2 applies a bisection method to search for the detection probability, which keeps a good detection performance. However, Active RIS 2 has a higher complexity than Active RIS 1, thus there exists a trade-off between the two active RIS algorithms.

%\begin{figure}\vspace{-20pt}
%	\centering
%	\includegraphics[width=2in]{fig5-1.eps}
%	\caption{The detection probability versus the sensing time.}
%	\label{fig5-1}%\vspace{-20pt}
%\end{figure}
%\begin{figure}\vspace{-5pt}
%	\centering
%	\includegraphics[width=2in]{fig6.eps}
%	\caption{The detection probability versus the transmit power of the PT.}
%	\label{fig6}\vspace{-15pt}
%\end{figure}

\begin{figure}[t]
	%	\vspace{-20pt}
	\centering
	\begin{minipage}[t]{0.24\textwidth}
		\centering
		\includegraphics[width=1.7in]{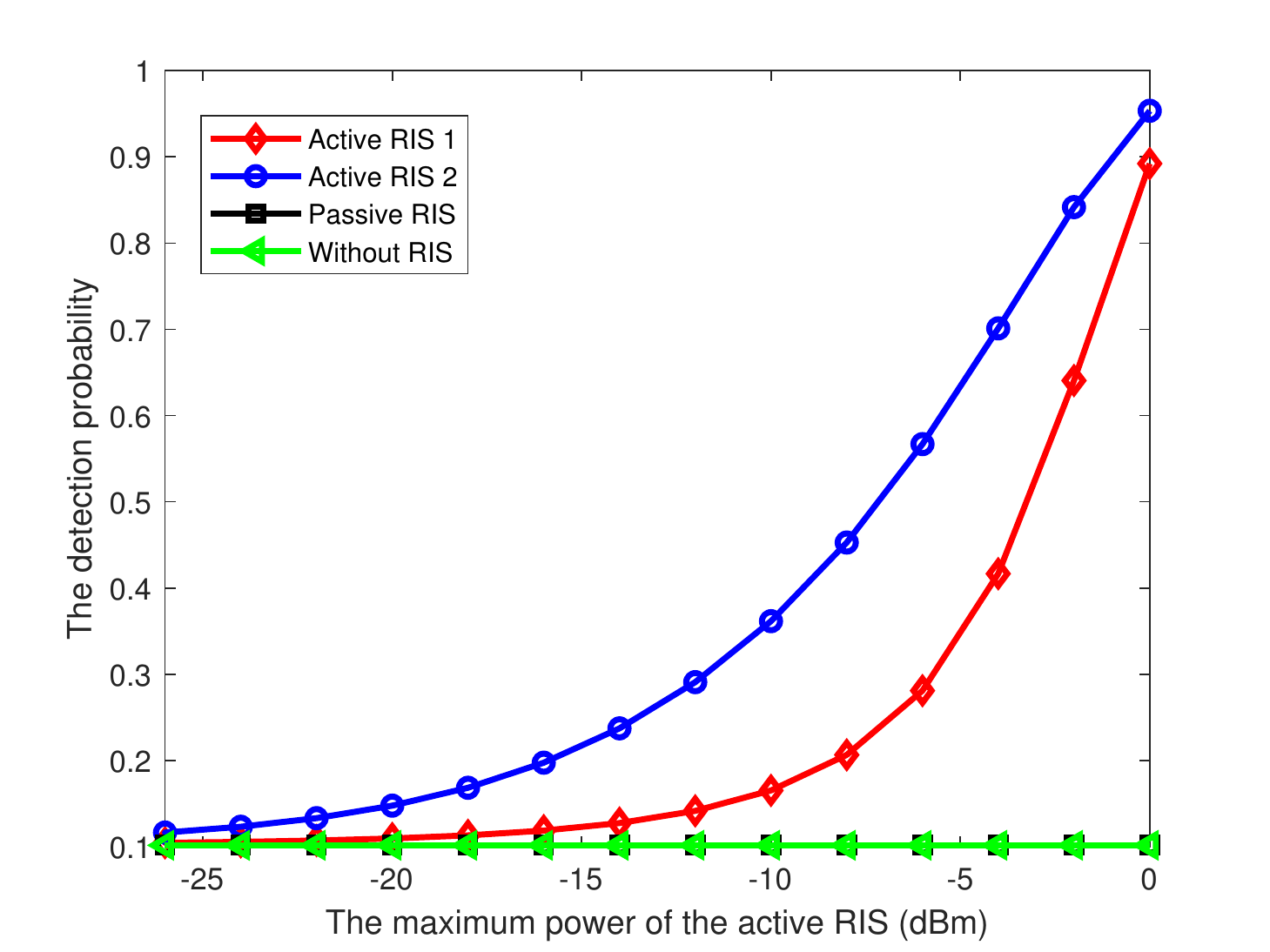}
		\caption{The detection probability versus the maximum power of the active RIS.}
		\label{fig7}%\vspace{-20pt}
	\end{minipage}
	\begin{minipage}[t]{0.24\textwidth}
		\centering
		\includegraphics[width=1.7in]{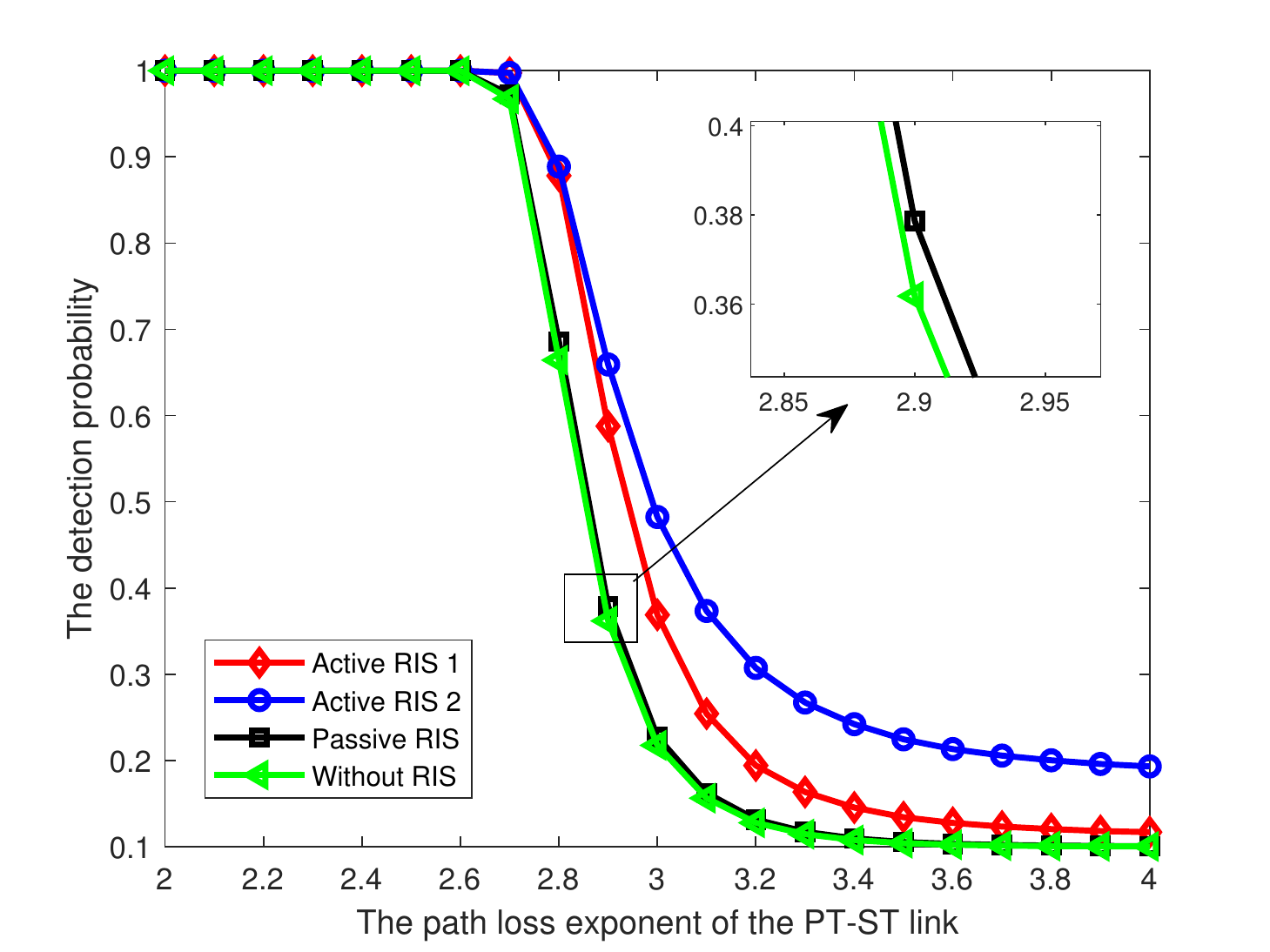}
		\caption{The detection probability versus the path loss exponent of the PT-ST link.}
		\label{fig10}%\vspace{-20pt}
	\end{minipage}
\end{figure}
\subsection{Performance Comparison of Proposed Algorithms}
Fig. \ref{fig6} gives the detection probability versus the transmit power of the PT under Passive RIS and Active RIS 1. From the figure, it can be observed that the detection probability under all algorithms increases with the increasing transmit power of the PT. This is due to the fact that the increase in transmit power of the PT leads to an increase in the signal strength transmitted to the ST, which makes the ST easier to detect the signal transmitted by the PT. The improved SNR at the ST facilitates the differentiation of the signal from background noise, thus enhancing the detection probability. Besides, we can find that with the same number of reflecting elements, the detection probability under active RIS is superior to passive RIS. The reasons behind this are twofold. First, from equations (\ref{eq9}) and (\ref{eq14}), it can be observed that active RIS, compared to passive RIS, not only adjusts the phase shift but also amplify signals, which can enhance the received signal strength at the ST. Second, different from active RIS-assisted wireless communications, where the thermal noise introduced by active RIS is one of the main bottlenecks of system performance, the thermal noise in active RIS-assisted spectrum sensing is beneficial for spectrum sensing. Furthermore, although the detection probability of active RIS outperforms that of passive RIS, passive RIS can achieve higher performance by increasing the number of reflecting elements. As shown in Proposition 6, when the number of reflecting elements in passive RIS is sufficiently large, its performance surpasses that of active RIS.

%\begin{figure}\vspace{-20pt}
%	\centering
%	\includegraphics[width=2in]{fig7.eps}
%	\caption{The detection probability versus the maximum power of the active RIS.}
%	\label{fig7}%\vspace{-20pt}
%\end{figure}
%\begin{figure}\vspace{-10pt}
%	\centering
%	\includegraphics[width=2in]{fig10.eps}
%	\caption{The detection probability versus the path loss exponent of the PT-ST link.}
%	\label{fig10}\vspace{-15pt}
%\end{figure}
Fig. \ref{fig7} shows the detection probability versus the maximum power of the active RIS under different algorithms. We can observe that the detection probability of the two algorithms under the active RIS increases with the increasing maximum power of the active RIS. This is due to the fact that the maximum power threshold of the active RIS expands the feasible region of the amplification factor. The amplification factor can be further increased to improve the power of the useful signal and the power of the thermal noise. Besides, we observe that under a larger power threshold of the active RIS, the detection probability under active RIS is significantly higher than {\color{black}that of passive RIS and without RIS}. This can be attributed to the additional power source provided by active RIS, which enhances the detection performance. Furthermore, we can also observe that under a larger power threshold of the active RIS, the detection probabilities of the two algorithms under active RIS gradually approach each other since the detection probabilities of both algorithms converge to 1.

Fig. \ref{fig10} shows the detection probability versus the path loss exponent under different algorithms. From the figure, it can be observed that as the path loss exponent increases, the detection probability decreases for all algorithms. This is because when the path loss exponent increases, the attenuation of the signal transmitted by the PT aggravates, resulting in a gradual decrease in the signal strength. As a result, the signal received at the ST becomes weaker, and compared to the background noise, the signal energy is lower, {\color{black}which makes it difficult to detect and distinguish the signal effectively}. Therefore, the detection probability decreases. Besides, under the same path loss exponent, the detection probability of active RIS is highest, followed by the passive RIS, and lowest in the case without RIS, which demonstrates the significant contribution of both passive and active RIS to spectrum sensing.

\begin{figure}%\vspace{-20pt}
	\centering
	\begin{subfigure}{0.49\linewidth}
		\centering
		\includegraphics[width=1\linewidth]{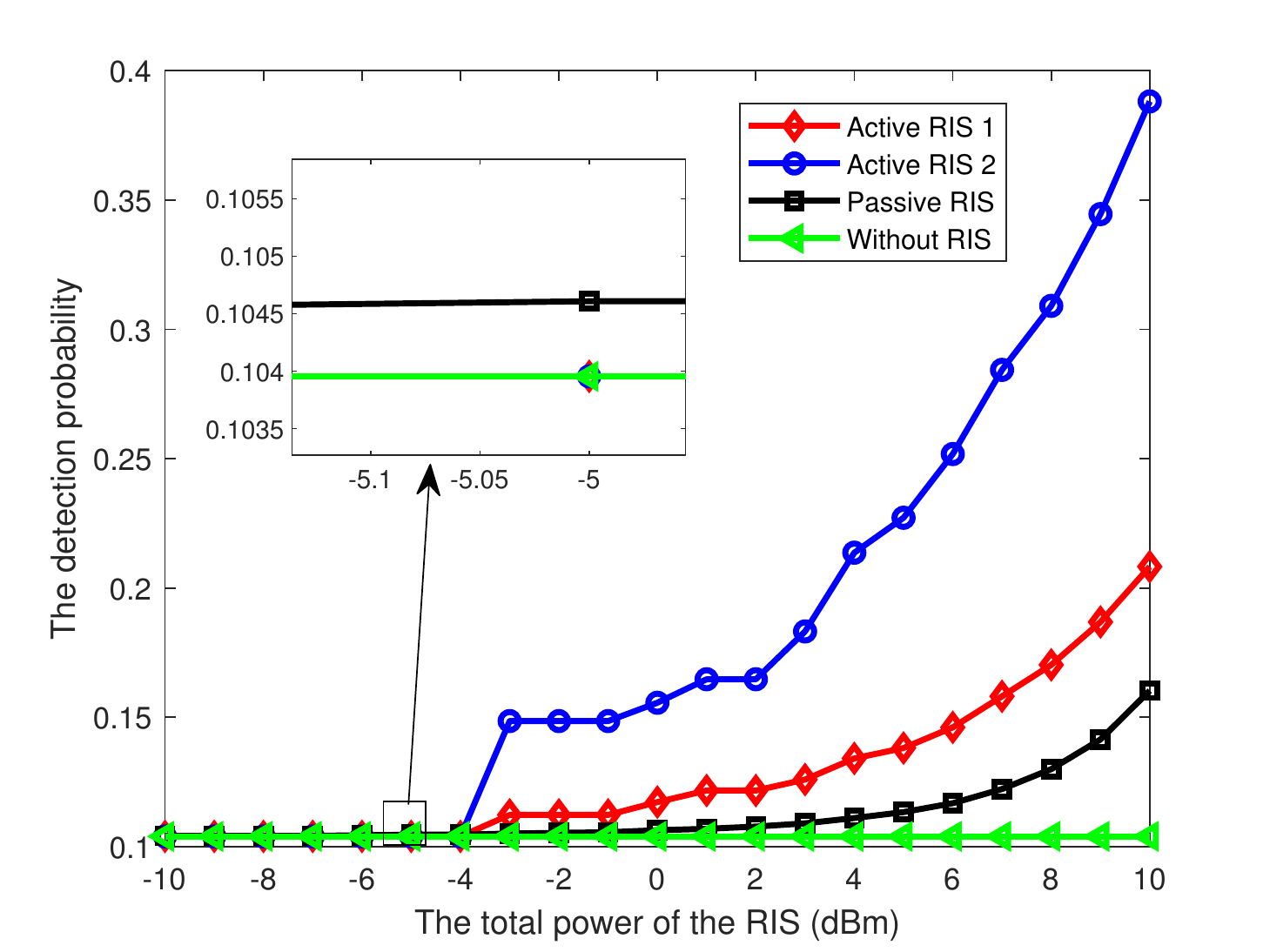}
		\caption{The detection probability versus the total power of the RIS.}
		\label{fig8}%文中引用该图片代号
	\end{subfigure}
	\centering
	\begin{subfigure}{0.49\linewidth}
		\centering
		\includegraphics[width=1\linewidth]{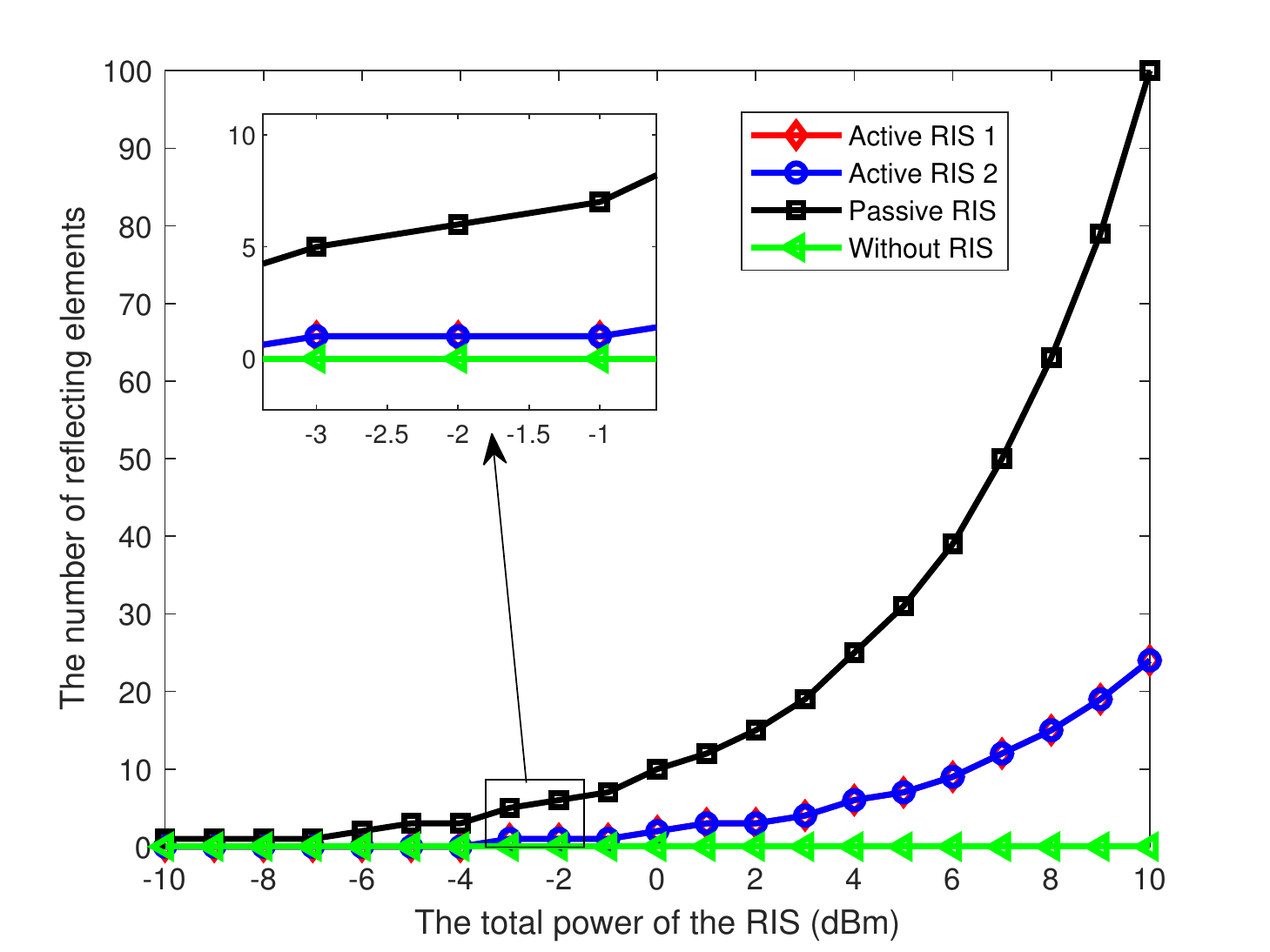}
		\caption{The number of reflecting elements versus the total power of the RIS.}
		\label{fig9}%文中引用该图片代号
	\end{subfigure}
	\caption{The impact of the total power of the RIS on the detection probability.}
	\label{fig89}
\end{figure}
\begin{figure}%\vspace{-15pt}
	\centering
	\includegraphics[width=2in]{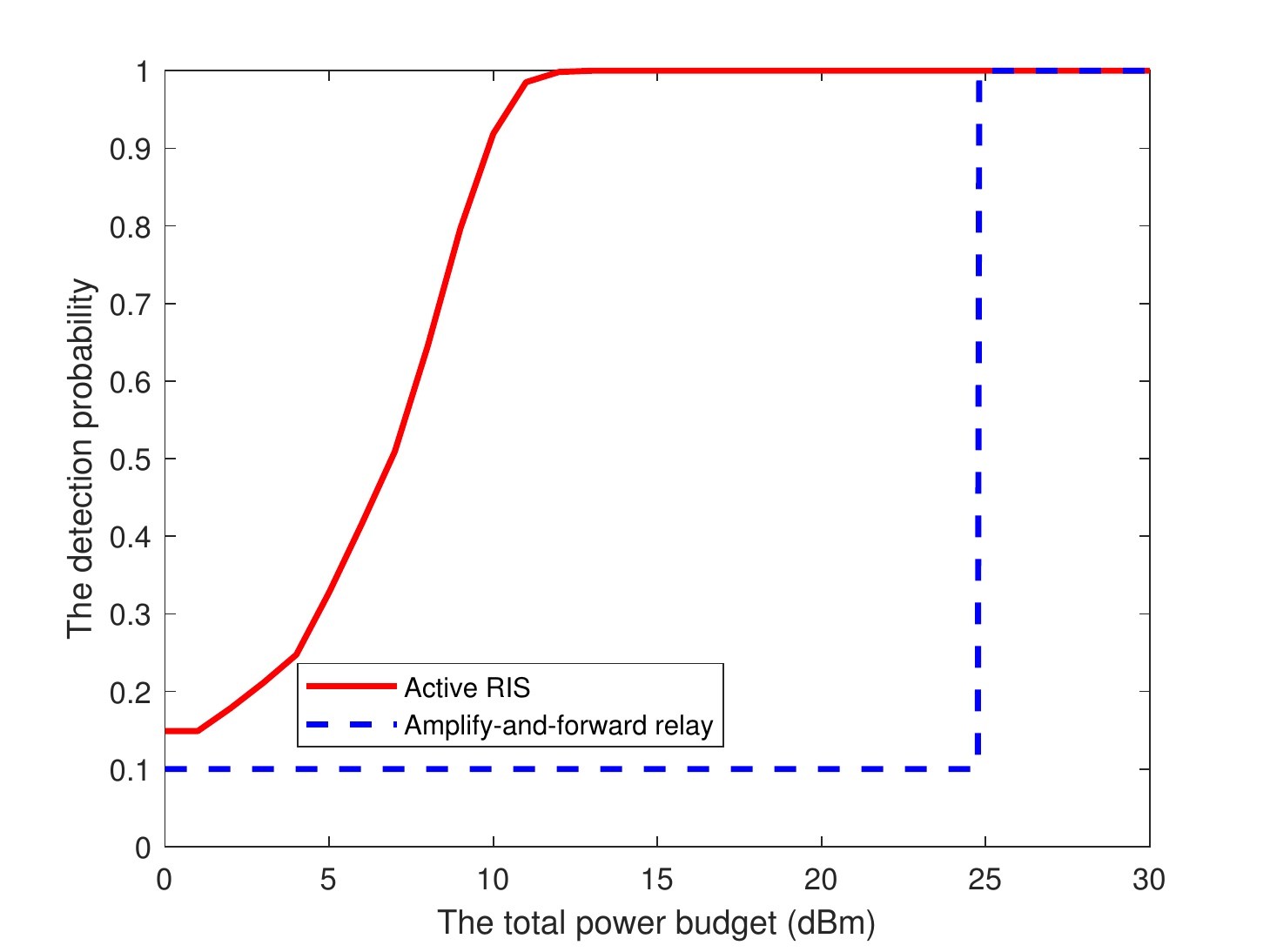}
	\caption{\color{black}The detection probability versus the total power budget.}
	\label{fig11}%\vspace{-15pt}
\end{figure}

Fig. \ref{fig89} shows the detection probability versus the total power of the RIS under different algorithms. We define the total power of the RIS as $P_{\rm T}$,  the power consumed by the phase shift switch and control circuit in each RIS element as $P_{\rm C}$, and the direct current biasing power used by the amplifier in each active element as $P_{\rm DC}$. {\color{black}We set $P_{\rm C}=-10$ dBm and $P_{\rm DC}=-5$ dBm\cite{i5}.} Then, under the same total power of the RIS, the number of reflecting elements for the passive RIS $N_{\rm pas}=\left\lfloor\frac{P_{\rm T}}{P_{\rm C}}\right\rfloor$ and the number of reflecting elements for the active RIS $N_{\rm act}=\left\lfloor\frac{P_{\rm T}-P_{\rm RIS}^{\max}}{P_{\rm C}+P_{\rm DC}}\right\rfloor$. From Fig. \ref{fig89}(a), we can observe that the detection probability under both active RIS and passive RIS increases with the increasing total power of the RIS. This is because the increase in the total power of the RIS supports the activation of more reflecting elements, as shown in Fig. \ref{fig89}(b), which provides a larger beamforming gain. Besides, at the low total power level of the RIS, the detection probability of passive RIS is higher than that of active RIS, and the detection probability of active RIS is equivalent to that of traditional without RIS. This is because when the total power of the RIS is low, the number of passive reflecting elements is usually larger than the number of active reflecting elements, and low total power may not be sufficient to activate the active reflecting elements, i.e., $N_{\rm act}=0$, as shown in Fig. \ref{fig89}(b). In this case, the performance of active RIS is similar to that without  RIS. However, at the high total power level of the RIS, despite the number of active reflecting elements is still less than that  of passive reflecting elements, the presence of amplification factors and amplified noise provides a higher detection probability under active RIS than under passive RIS. {\color{black}We can observe from Fig. \ref{fig89}(b) that the curves for active RIS 1 and active RIS 2 overlap. This is because under the same total power budget of the RIS, both sensing algorithms for active RIS have the same number of available reflecting elements. Despite having the same number of reflecting elements, their sensing performance is different, as shown in Fig. \ref{fig89}(a).}

\color{black}
Fig. \ref{fig11} gives the detection probability versus the total power budget under the active RIS and full-duplex amplify-and-forward relay, where $P_{\rm RIS}^{\max}= -10$ dBm. We consider the power consumption model of the amplify-and-forward relay in \cite{j1}, the number of transmitting and receiving antenna are 5, the power dissipation at each transmitting and receiving antenna are 14 dBm, and the power consumption of self-interference cancellation at the relay is 17 dBm. To focus on the comparison, the direct link from the PT to the ST is ignored, and perfect self-interference cancellation is assumed for the full-duplex amplify-and-forward relay. With the increase in the total power budget, the detection probabilities of both spectrum sensing algorithms will undoubtedly improve. Furthermore, we can observe that active RIS-aided spectrum sensing can achieve a detection probability close to 1 under a low power budget, while amplify-and-forward relay requires a larger power budget. There are two reasons behind this: First,  the power consumption of the active RIS is lower compared to the relay. Under the same power consumption, the active RIS can employ more reflecting elements, achieving higher passive beamforming gains. Second, the active RIS not only amplifies the signal but also adjusts the phase shift, thus enhancing the received SINR. This phenomenon also indicates that the active RIS is more energy-efficient compared to the amplify-and-forward relay.
\color{black}

%\begin{figure}
%	\centering
%	\includegraphics[width=3in]{fig8.eps}
%	\caption{The detection probability versus the total power of the RIS.}
%	\label{fig0}%\vspace{-20pt}
%\end{figure}
%\begin{figure}
%	\centering
%	\includegraphics[width=3in]{fig9.eps}
%	\caption{The number of reflecting elements versus the total power of the RIS..}
%	\label{fig0}%\vspace{-20pt}
%\end{figure}

\vspace{-5pt}
\section{Conclusion}
This paper analyzes the impact of the RIS on spectrum sensing. In particular, the optimization problems are respectively formulated for the passive RIS and the active RIS with the objective of maximizing detection probability while satisfying constraints on the maximum power of the active RIS and the maximum probability of false alarm. Due to the intractable problems, we propose a one-stage optimization algorithm with inner approximation and a two-stage optimization algorithm with a bisection method to deal with the highly non-convex problems.  Moreover, we explore the number configuration for the passive RIS and the active RIS, analyze how many reflecting elements are needed to achieve the detection probability close to 1, and compare the detection performance for the passive RIS and the active RIS. Simulation results demonstrate that the proposed algorithms outperform baseline algorithms under the same parameter configuration.

\appendices
\section{}
{\color{black}Based on the test statistic} for the energy detector, we have $T(y){=}\frac{1}{I}\sum_{i=1}^I|y(i)|^2$. According to the CLT, when $I$ is large, the PDF of  $T(y)$ can be approximated as a Gaussian distribution, where the mean $u_0^{\rm pas}=\mathbb{E}\{\frac{1}{I}\sum_{i=1}^I|y(i)|^2\}\approx \mathbb{E}\{|y(i)|^2\}=\mathbb{D}\{\boldsymbol{\rm w}^H\boldsymbol{\rm n}(i)\}+\mathbb{E}^2\{\boldsymbol{\rm w}^H\boldsymbol{\rm n}(i)\}=\delta^2+0=\delta^2$. The variance $v_0^{\rm pas}$ is 
\begin{equation}\label{eq87}
\begin{split}
&v_0^{\rm pas}{=}\mathbb{E}\{|\frac{1}{I}\sum_{i=1}^I|y(i)|^2{-}u_0^{\rm pas}|^2\}\\
&{\approx}\frac{1}{I}\mathbb{E}\{(|\boldsymbol{\rm w}^H\boldsymbol{\rm n}(i)|^2{-}u_0^{\rm pas})^2\}\\
&{=}\frac{1}{I}(\mathbb{E}\{|\boldsymbol{\rm w}^H\boldsymbol{\rm n}(i)|^4\}{-}2u_0^{\rm pas}\mathbb{E}\{|\boldsymbol{\rm w}^H\boldsymbol{\rm n}(i)|^2\}{+}(u_0^{\rm pas})^2)\\
&{=}\frac{1}{I}(\mathbb{E}\{|\boldsymbol{\rm w}^H\boldsymbol{\rm n}(i)|^4\}{-}(u_0^{\rm pas})^2){=}\frac{1}{I}(\mathbb{E}\{|\boldsymbol{\rm w}^H\boldsymbol{\rm n}(i)|^4\}{-}\delta^4).
\end{split}
\end{equation}
We assume that the received signal $\boldsymbol{\rm w}^H\boldsymbol{\rm n}(i)$ is CSCG. In this case, $\mathbb{E}\{|\boldsymbol{\rm w}^H\boldsymbol{\rm n}(i)|^4\}=2\delta^4$. Substituting it into (\ref{eq87}), we can obtain $v_0^{\rm pas}=\frac{\delta^4}{I}$.

The proof is completed.

\section{}
From (\ref{eq3}), under hypothesis $\mathcal{H}_1$, both $s(i)$ and $\boldsymbol{\rm n}(i)$ are CSCG, we have
\begin{equation}\label{eq88}
\begin{split}
u_1^{\rm pas}&=\mathbb{E}\{\frac{1}{I}\sum_{i=1}^I|y(i)|^2\}\approx \mathbb{E}\{|y(i)|^2\}\\
&=\mathbb{E}\{|\boldsymbol{\rm w}^H(\boldsymbol{{\rm h}}_{\rm d}+\boldsymbol{\rm H}\boldsymbol{\rm \Theta}\boldsymbol{\rm h}_{{\rm r}})\sqrt{p}s(i)+\boldsymbol{\rm w}^H \boldsymbol{\rm n}(i)|^2\}\\
&=\mathbb{D}\{\boldsymbol{\rm w}^H(\boldsymbol{{\rm h}}_{\rm d}+\boldsymbol{\rm H}\boldsymbol{\rm \Theta}\boldsymbol{\rm h}_{{\rm r}})\sqrt{p}s(i)+\boldsymbol{\rm w}^H \boldsymbol{\rm n}(i)\}\\
&+\mathbb{E}^2\{\boldsymbol{\rm w}^H(\boldsymbol{{\rm h}}_{\rm d}+\boldsymbol{\rm H}\boldsymbol{\rm \Theta}\boldsymbol{\rm h}_{{\rm r}})\sqrt{p}s(i)+\boldsymbol{\rm w}^H \boldsymbol{\rm n}(i)\}\\
&=p|\boldsymbol{\rm w}^H(\boldsymbol{{\rm h}}_{\rm d}+\boldsymbol{\rm H}\boldsymbol{\rm \Theta}\boldsymbol{\rm h}_{{\rm r}})|^2+\delta^2.
\end{split}
\end{equation}

Furthermore, based on the fact that $s(i)$ and $\boldsymbol{\rm n}(i)$ are independent and with zero mean, we can obtain the following variance as (\ref{eq89}). Suppose that the received signal $\boldsymbol{\rm w}^H(\boldsymbol{{\rm h}}_{\rm d}+\boldsymbol{\rm H}\boldsymbol{\rm \Theta}\boldsymbol{\rm h}_{{\rm r}})\sqrt{p}s(i)$ is CSCG, then we have $\mathbb{E}\{|\boldsymbol{\rm w}^H(\boldsymbol{{\rm h}}_{\rm d}+\boldsymbol{\rm H}\boldsymbol{\rm \Theta}\boldsymbol{\rm h}_{{\rm r}})\sqrt{p}s(i)|^4\}=2p^2|\boldsymbol{\rm w}^H(\boldsymbol{{\rm h}}_{\rm d}+\boldsymbol{\rm H}\boldsymbol{\rm \Theta}\boldsymbol{\rm h}_{{\rm r}})|^4$. By substituting it into (\ref{eq89}), we obtain $v_1^{\rm pas}=\frac{1}{I}(p^2|\boldsymbol{\rm w}^H(\boldsymbol{{\rm h}}_{\rm d}+\boldsymbol{\rm H}\boldsymbol{\rm \Theta}\boldsymbol{\rm h}_{{\rm r}})|^4+\delta^4
+2\delta^2p|\boldsymbol{\rm w}^H(\boldsymbol{{\rm h}}_{\rm d}+\boldsymbol{\rm H}\boldsymbol{\rm \Theta}\boldsymbol{\rm h}_{{\rm r}})|^2)=\frac{1}{I}(\gamma^{\rm pas}+1)^2\delta^4$, where $\gamma^{\rm pas}=\frac{p|\boldsymbol{\rm w}^H(\boldsymbol{{\rm h}}_{\rm d}+\boldsymbol{\rm H}\boldsymbol{\rm \Theta}\boldsymbol{\rm h}_{{\rm r}})|^2}{\delta^2}$.
\begin{figure*}\vspace{-30pt}
	\begin{equation}\label{eq89}
	\begin{split}
	v_1^{\rm pas}&{=}\mathbb{E}\{|\frac{1}{I}\sum_{i=1}^I|y(i)|^2{-}u_1^{\rm pas}|^2\}\approx\mathbb{E}\{(\frac{1}{I}\sum_{i=1}^I|\boldsymbol{\rm w}^H(\boldsymbol{{\rm h}}_{\rm d}{+}\boldsymbol{\rm H}\boldsymbol{\rm \Theta}\boldsymbol{\rm h}_{{\rm r}})\sqrt{p}s(i){+}\boldsymbol{\rm w}^H \boldsymbol{\rm n}(i)|^2{-}u_1^{\rm pas})^2\}\\
	&{=}\frac{1}{I}\mathbb{E}\{(|\boldsymbol{\rm w}^H(\boldsymbol{{\rm h}}_{\rm d}{+}\boldsymbol{\rm H}\boldsymbol{\rm \Theta}\boldsymbol{\rm h}_{{\rm r}})\sqrt{p}s(i){+}\boldsymbol{\rm w}^H \boldsymbol{\rm n}(i)|^2{-}(p|\boldsymbol{\rm w}^H(\boldsymbol{{\rm h}}_{\rm d}{+}\boldsymbol{\rm H}\boldsymbol{\rm \Theta}\boldsymbol{\rm h}_{{\rm r}})|^2{+}\delta^2))^2\}{=}\frac{1}{I}\mathbb{E}\{(|\boldsymbol{\rm w}^H(\boldsymbol{{\rm h}}_{\rm d}{+}\boldsymbol{\rm H}\boldsymbol{\rm \Theta}\boldsymbol{\rm h}_{{\rm r}})\sqrt{p}s(i)|^2\\
	&{+}(\boldsymbol{\rm w}^H(\boldsymbol{{\rm h}}_{\rm d}{+}\boldsymbol{\rm H}\boldsymbol{\rm \Theta}\boldsymbol{\rm h}_{{\rm r}})\sqrt{p}s(i))(\boldsymbol{\rm w}^H \boldsymbol{\rm n}(i))^H{+}(\boldsymbol{\rm w}^H(\boldsymbol{{\rm h}}_{\rm d}{+}\boldsymbol{\rm H}\boldsymbol{\rm \Theta}\boldsymbol{\rm h}_{{\rm r}})\sqrt{p}s(i))^H\boldsymbol{\rm w}^H \boldsymbol{\rm n}(i){+}|\boldsymbol{\rm w}^H \boldsymbol{\rm n}(i)|^2{-}(p|\boldsymbol{\rm w}^H(\boldsymbol{{\rm h}}_{\rm d}{+}\boldsymbol{\rm H}\boldsymbol{\rm \Theta}\boldsymbol{\rm h}_{{\rm r}})|^2\\
	&{+}\delta^2))^2\}{=}\frac{1}{I}(\mathbb{E}\{|\boldsymbol{\rm w}^H(\boldsymbol{{\rm h}}_{\rm d}{+}\boldsymbol{\rm H}\boldsymbol{\rm \Theta}\boldsymbol{\rm h}_{{\rm r}})\sqrt{p}s(i)|^4\}{+}\mathbb{E}\{|\boldsymbol{\rm w}^H \boldsymbol{\rm n}(i)|^4\}{-}(p|\boldsymbol{\rm w}^H(\boldsymbol{{\rm h}}_{\rm d}{+}\boldsymbol{\rm H}\boldsymbol{\rm \Theta}\boldsymbol{\rm h}_{{\rm r}})|^2{-}\delta^2)^2).\\
	\end{split}
	\end{equation}
		\hrulefill
\end{figure*}

The proof is completed.

\section{}
Let us first recall Proposition 2, when the number of samples is sufficiently large, for the passive RIS, the PDF of $T(y)$ under hypothesis $\mathcal{H}_1$ can be approximated by a Gaussian distribution with the mean $u_1^{\rm pas}$ and variance $v_1^{\rm pas}$. Then, we have
\begin{equation}\label{eq90}
\begin{split}
P_d^{\rm pas}={\rm Pr}(T(y)>\epsilon|\mathcal{H}_1)=\int_{\epsilon}^{\infty}f_{T>\epsilon|\mathcal{H}_1}(x)dx,
\end{split}
\end{equation}
where
$f_{T>\epsilon|\mathcal{H}_1}(x)=\frac{1}{\sqrt{2\pi v_1^{\rm pas}}} \exp(-\frac{(x-u_1^{\rm pas})^2}{2v_1^{\rm pas}})$.
We need to simplify the above equations to the complementary distribution function of the standard Gaussian, i.e., $Q(x)=\frac{1}{\sqrt{2\pi}}\int_{x}^{\infty} \exp(-\frac{t^2}{2})\, dt$. Then, let us define $t=\frac{x-u_1^{\rm pas}}{\sqrt{v_1^{\rm pas}}}$, then we have $t=\frac{\epsilon-u_1^{\rm pas}}{\sqrt{v_1^{\rm pas}}}$ and $dx=\sqrt{v_1^{\rm pas}}dt$. Thus, we have
\begin{equation}\label{eq92}
\begin{split}
P_d^{\rm pas}=\frac{1}{\sqrt{2\pi }}\int_{\frac{\epsilon-u_1^{\rm pas}}{\sqrt{v_1^{\rm pas}}}}^{\infty} \exp(-\frac{t^2}{2})dt=Q(\frac{\epsilon-u_1^{\rm pas}}{\sqrt{v_1^{\rm pas}}}).
\end{split}
\end{equation}
Then, substituting $u_1^{\rm pas}$ and $v_1^{\rm pas}$ into (\ref{eq92}), we have
\begin{equation}\label{eq93}
\begin{array}{l}
P_d^{\rm pas}=Q(\frac{\epsilon-(p|\boldsymbol{\rm w}^H(\boldsymbol{{\rm h}}_{\rm d}+\boldsymbol{\rm H}\boldsymbol{\rm \Lambda}\boldsymbol{\rm \Theta}\boldsymbol{\rm h}_{{\rm r}})|^2+\delta^2)}{\sqrt{\frac{1}{I}(\gamma^{\rm pas}+1)^2\delta^4}})\\
~~~~~~=Q((\frac{\epsilon}{\delta^2}-\gamma^{\rm pas}-1)\sqrt{\frac{I}{(1+\gamma^{\rm pas})^2}}).
\end{array}
\end{equation}

The proof is completed.

\section{}
Define $\boldsymbol{\rm x}=[x_1,\cdots,x_n,\cdots,x_N]^H$, where $x_n=e^{j\theta_n}$. Then, we have
\begin{equation}\label{eq59}
\begin{split}
&{\rm Pr}(\mathcal{H}_1)(p\|\boldsymbol{\rm \Lambda}\boldsymbol{\rm \Theta}\boldsymbol{\rm h}_{\rm r}\|^2+\sigma^2\|\boldsymbol{\rm \Lambda}\boldsymbol{\rm \Theta}\|_F^2)\\
&=\boldsymbol{\rm x}^H\left({\rm Pr}(\mathcal{H}_1)(p{\rm diag}(\boldsymbol{\rm h}_{\rm r}^H)\boldsymbol{\rm \Lambda}^H\boldsymbol{\rm \Lambda}{\rm diag}(\boldsymbol{\rm h}_{\rm r})+\sigma^2\boldsymbol{\rm \Lambda}^H\boldsymbol{\rm \Lambda})\right)\boldsymbol{\rm x}\\
&=\boldsymbol{\rm x}^H\boldsymbol{\rm C}\boldsymbol{\rm x}\overset{(a)}{=}N\frac{\boldsymbol{\rm x}^H\boldsymbol{\rm C}\boldsymbol{\rm x}}{\boldsymbol{\rm x}^H\boldsymbol{\rm x}},
\end{split}
\end{equation}
where $(a)$ utilizes $\boldsymbol{\rm x}^H\boldsymbol{\rm x}=N$. It is worth noting that $\frac{\boldsymbol{\rm x}^H\boldsymbol{\rm C}\boldsymbol{\rm x}}{\boldsymbol{\rm x}^H\boldsymbol{\rm x}}$ is the form of Rayleigh quotient, where $\boldsymbol{\rm C}$ is a Hermitian matrix. Since $\boldsymbol{\rm C}$ is a symmetric matrix, we can apply eigenvalue decomposition $\boldsymbol{\rm C}=\boldsymbol{\rm P}\boldsymbol{\rm \Sigma}\boldsymbol{\rm P}^H$, where $\boldsymbol{\rm \Sigma}={\rm diag}(\lambda_1,\lambda_2,\cdots,\lambda_N)$ is diagonal eigenvalue matrix and $\lambda_1<\lambda_2<\cdots<\lambda_N$, $\boldsymbol{\rm P}$ is matrix with respect to eigenvectors and $\boldsymbol{\rm P}\boldsymbol{\rm P}^H=\boldsymbol{\rm I}$. Defining $\boldsymbol{\rm z}=\boldsymbol{\rm P}^H\boldsymbol{\rm x}$, we have
\begin{equation}\label{eq60}
\begin{split}
\frac{\boldsymbol{\rm x}^H\boldsymbol{\rm C}\boldsymbol{\rm x}}{\boldsymbol{\rm x}^H\boldsymbol{\rm x}}=\frac{\boldsymbol{\rm x}^H\boldsymbol{\rm P}\boldsymbol{\rm \Sigma}\boldsymbol{\rm P}^H\boldsymbol{\rm x}}{\boldsymbol{\rm x}^H\boldsymbol{\rm P}\boldsymbol{\rm P}^H\boldsymbol{\rm x}}=\frac{\boldsymbol{\rm z}^H\boldsymbol{\rm \Sigma}\boldsymbol{\rm z}}{\boldsymbol{\rm z}^H\boldsymbol{\rm z}}=\frac{\sum\limits_{n=1}^N\lambda_nz_n^2}{\sum\limits_{n=1}^Nz_n^2}.
\end{split}
\end{equation}
Then, we have
\begin{equation}\label{eq61}
\begin{split}
\lambda_1=\frac{\sum\limits_{n=1}^N\lambda_1z_n^2}{\sum\limits_{n=1}^Nz_n^2}\leq \frac{\sum\limits_{n=1}^N\lambda_nz_n^2}{\sum\limits_{n=1}^Nz_n^2} \leq \frac{\sum\limits_{n=1}^N\lambda_Nz_n^2}{\sum\limits_{n=1}^Nz_n^2}=\lambda_N.
\end{split}
\end{equation}
Since ${\rm Pr}(\mathcal{H}_1)p{\rm diag}(\boldsymbol{\rm h}_{\rm r}^H)\boldsymbol{\rm \Lambda}^H\boldsymbol{\rm \Lambda}{\rm diag}(\boldsymbol{\rm h}_{\rm r})+\sigma^2\boldsymbol{\rm \Lambda}^H\boldsymbol{\rm \Lambda})$ is a symmetric matrix, its eigenvalues are diagonal elements. Thus, $\lambda_{\min}$ and $\lambda_{\max}$ are the smallest and largest eigenvalues respectively, and are also the smallest and largest elements of the principal diagonal.

The proof is completed.

\end{document}